\documentclass[a4paper,fleqn,usenatbib]{mnras}

\usepackage{newtxtext,newtxmath}
\usepackage[T1]{fontenc}
\usepackage{ae,aecompl}

\usepackage{graphicx}	% Including figure files
\usepackage{amsmath}	% Advanced maths commands

\usepackage{amssymb}	% Extra maths symbols
\usepackage[english]{babel} %remove after writing
\usepackage{blindtext} %remove after writing
\usepackage{lipsum} %remove after writing

\title[On the environmental distribution of S0 galaxies]{Connecting galaxy structure and star formation: the role of environment in formation of S0 galaxies}

\author[P. K. Mishra et al.]{
Preetish K. Mishra,$^{1}$\thanks{E-mail: preetish@ncra.tifr.res.in}
Yogesh Wadadekar,$^{1}$\thanks{E-mail: yogesh@ncra.tifr.res.in}
Sudhanshu Barway,$^{2}$\thanks{E-mail: sudhanshu.barway@iiap.res.in  }
\\
$^{1}$National Centre for Radio Astrophysics, TIFR, Post Bag 3, Ganeshkhind, Pune 411007, India\\
$^{2}$Indian Institute of Astrophysics, Koramangala II Block, Bengaluru 560034, India
}
\date{Accepted XXX. Received YYY; in original form ZZZ}

\pubyear{2018}

\begin{document}
\label{firstpage}
\pagerange{\pageref{firstpage}--\pageref{lastpage}}
\maketitle

\begin{abstract}

In this work, we investigate the reason behind the increased occurrence
of S0 galaxies in high density environments. Our sample comprises of
$\sim$ 2500 spiral and $\sim$ 2000 S0 galaxies spanning a wide range of
environments. Dividing the galaxies into categories of classical and
pseudobulge hosting spiral and S0 galaxies, we have studied their
properties as a function of the environment. We find that the fraction
of pseudobulge hosting disc galaxies decreases with increase in
density. The classical bulge hosting spirals and S0 galaxies follow a
similar trend in less dense environments but towards higher
densities, we observe an increase in the fraction of classical bulge host
S0 galaxies at the expense of spirals. Comparing the structural
and the star formation properties of galaxies on the size-mass and
$NUV-r$ colour-mass planes respectively, we infer that classical bulge
hosting spirals are likely to get transformed into S0
morphology. We notice a trend of galaxy structure with environment such that the fraction of classical bulge hosting spiral galaxies is found to increase with environment density. We also find that among classical bulge hosting spirals, the fraction of quenched galaxies increases in denser environments. We surmise that the existence of more classical bulge hosting spirals galaxies and more efficient quenching leads to the observed increased occurrence of S0 galaxies in high density environments. The relation between galaxy structure and environment also exists for the disc galaxies irrespective of their visual morphology, which is driven mainly by halo mass.

\end{abstract}

\begin{keywords}
galaxies: bulges -- galaxies: formation -- galaxies: evolution
\end{keywords}

%%%%%%%%%%%%%%%%%%%%%%%%%%%%%%%%%%%%%%%%%%%%%%%%%%

%%%%%%%%%%%%%%%%% BODY OF PAPER %%%%%%%%%%%%%%%%%%

\section{Introduction}

\label{sec:intro}

A fundamental question regarding galaxy evolution is what causes the galaxies to acquire their current visual appearance or morphology since their formation. In the standard picture of galaxy formation in the $\Lambda$CDM universe, galaxies are formed inside dark matter haloes through the cosmological infall of gas which then settles and forms stars \citep{White1978, Benson2010}. Ever since their formation at high redshift,  galaxies have undergone a major evolution in their properties and, now form a inhomogeneous population of objects with respect to their morphology \citep{Concelise2014}. In order to systematically study galaxies, \cite{Hubble1936} classified galaxies based on their visual appearance on the tuning fork diagram. The three major morphological classes in this scheme are the ellipticals, S0 and spiral galaxies which are different from each other in their structural and star formation properties (\cite{Buta2013} and references therein). These differences of properties among the morphological classes are thought to arise mainly from combination of two factors: the different initial conditions for progenitors (or, difference in nature) of different morphological classes and the different interaction with environment  (different nurture scenario) which the progenitors undergo along their evolutionary path. To what extent the galaxy properties are shaped by the nature and nurture scenarios, is a matter of debate \citep{Irwin1995}. It is entirely possible that there are different answers to this question for different morphological classes but it is important to understand how the galaxies came to exist in their current morphology. One possible approach can be to improve our knowledge on one morphological class at a time and then integrate the understanding into a big picture. In this work, we have tried to gain new insights on the formation and the distribution of S0 galaxies in different environments.\par

The S0 galaxies have been placed in-between the ellipticals and the spirals on the Hubble tuning fork diagram \citep{Hubble1936}. They are similar to spiral galaxies in their structure, having no (or very faint) spiral arms in their discs, but are similar to the ellipticals in their colours. The formation of S0 galaxies is an active area of research and there has been major progress which has shaped our understanding of S0 galaxies. The S0 galaxies are thought to be transformed spirals \citep{Salamanca2006, Barway2009, Laurikainen2010}. In the literature, one finds a number of mechanisms for this morphological transformation. One way of converting a spiral into an S0 galaxy is through merger with other galaxies. Past studies have shown that a spiral galaxy undergoing a major merger with similarly massive companion, or undergoing a series of minor merger with smaller (in mass ratio 7:1) companion galaxy can lose its spiral arms while still retaining their disc structure (\citep{Querejeta2015,Tapia2017} and references therein). The other significant channel of making an S0 galaxy from a spiral is via shut down or quenching of star formation which leads to disc fading and disappearance of spiral arms \citep{Bekki2002, Barway2007, Barway2009, Bekki2011,Rizzo2018}. This quenching can happen due to internal processes acting from within the galaxy or due to environmental effects. Some of the internal processes responsible for quenching are: feedback due to supernovae or the central AGN, morphological quenching etc \citep{Cox2006,Martig2009}. The environmental quenching processes include star formation shutdown, removal of disc gas due to ram pressure stripping  in cluster environments, tidal interaction and galaxy harassment, heating of infalling gas by the dark matter halo etc. \citep{Gunn1972,Larson1980, Moore1996, Peng2015}. All these processes can potentially transform spiral galaxies into S0 galaxies but the extent to which they are important seems to depend on galaxy mass \citep{Fraser2018} and the environment in which the progenitor spiral lives. \par 

Observationally, it has been seen that S0 galaxies are more commonly found in high density environments. Previous studies have shown that the fraction of S0 galaxies increases with increase in local environmental density which happens along with a decrease in spiral galaxy fraction as the density increases \citep{Dressler1980,Postman1984}. This trend of galaxy morphology with local density is known to exist in scales of galaxy groups, clusters and superclusters \citep{Giovanelli1986,Dressler1997,Fasano2000,Wilman2009}. The morphology density relation is known to exist since $z\sim1$\citep{Postman2005} although its form has changed due to increase in the fraction of early type (Ellipticals+S0) galaxy fraction through the morphological transformation of spirals in denser environments \citep{Smith2005, Park2007, Houghton2015}.\par

These studies, while qualitatively explaining the trends of disc galaxy morphology with the environment, often ignore the possibility of a biased environmental distribution of progenitor spirals. Morphology, when defined visually, results from a combination of smoothness of light distribution (which depends on distribution star forming regions, dust lanes etc.) and dynamical structure of the galaxy (presence of disc and spheroids)\citep{Sandage1961,Wel2010}. In this sense, morphology already has a connection with star formation which depends strongly on environment. This interdependence of morphology, star formation and environment makes it difficult to gain insights on the nature and nurture aspects of morphological transformation. One can gain much more physical insight on the origin of the morphology density relation when only structural parameters are used to define the galaxy morphology \citep{Wel2010}. \par

Instead of relying totally on visual morphology, one can take an approach in which a galaxy is thought to be composed of a number of distinct dynamical components. For example, one can think of disc galaxies as a minimal combination of a bulge and a disc component. While discs are the defining physical components in disc galaxies, the bulges in disc galaxies might differ in their structural properties. Previous studies have shown that bulges of disc galaxies are of two major kinds. The first kind of bulges are the classical bulges which are featureless, spheroidal in shape and have kinematically hot, old stellar population \citep{Kormendy2004,Fisher2016}. They are structurally similar to elliptical galaxies and are known to follow the same scaling relations in the fundamental plane of galaxies as ellipticals. Classical bulges are formed in violent processes like galaxy mergers or sinking and coalescence of giant gas clumps in high redshift galaxy discs \citep{Elmegreen2008, Kormendy2016}. The second kind of bulges are known as pseudobulges which have disc like visual and kinematic structure \citep{Kormendy2004,Fisher2016}. They are thought to form secularly due to slow rearrangement of material in the disc by some internal dynamical structure such as bars, ovals etc \citep{Eliche-Moral2011, Kormendy2016}.\par

\begin{figure*}
%\centering
\includegraphics[width=.35\textwidth]{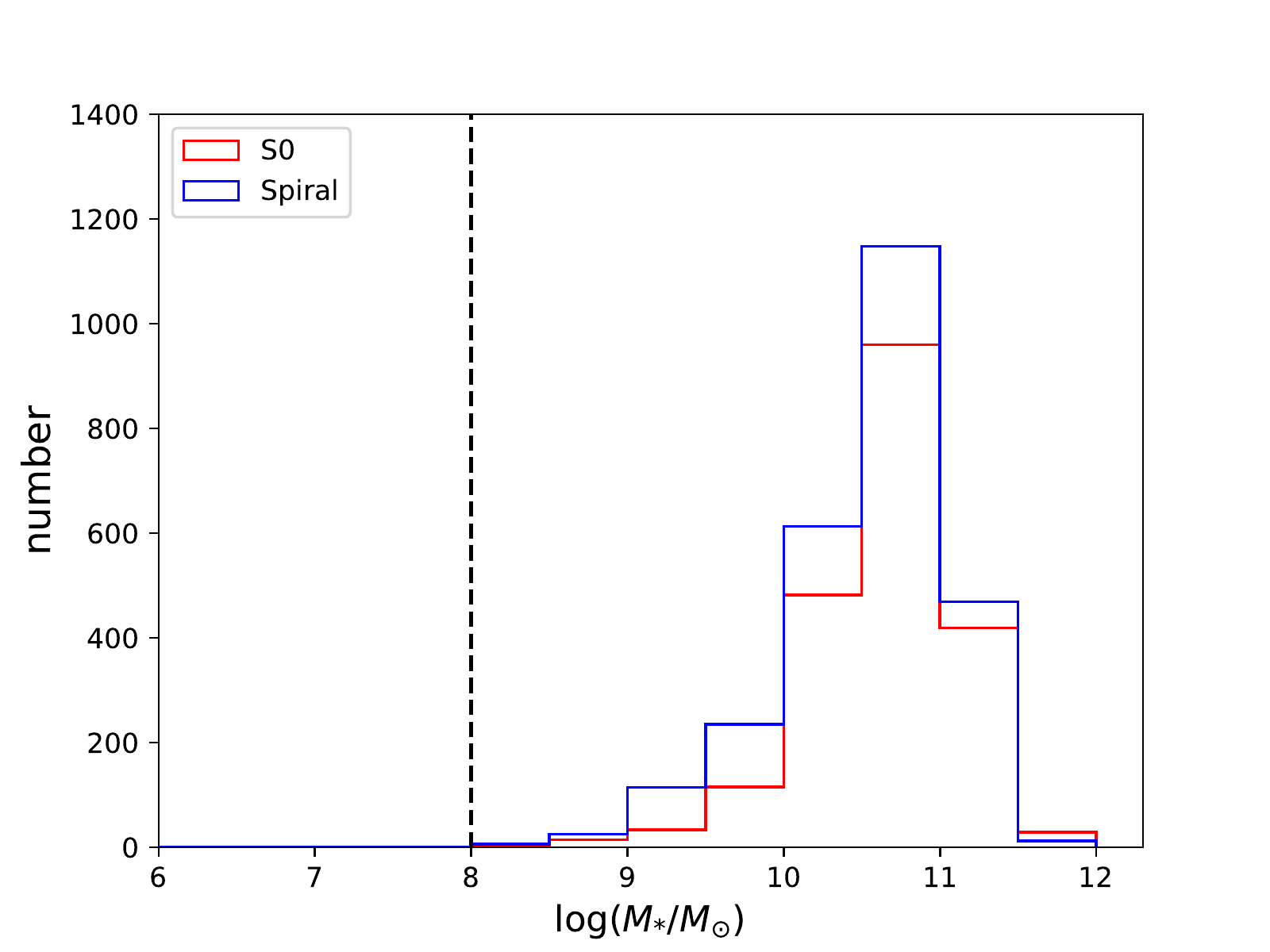}\hspace{-1.8em}
\includegraphics[width=.35\textwidth]{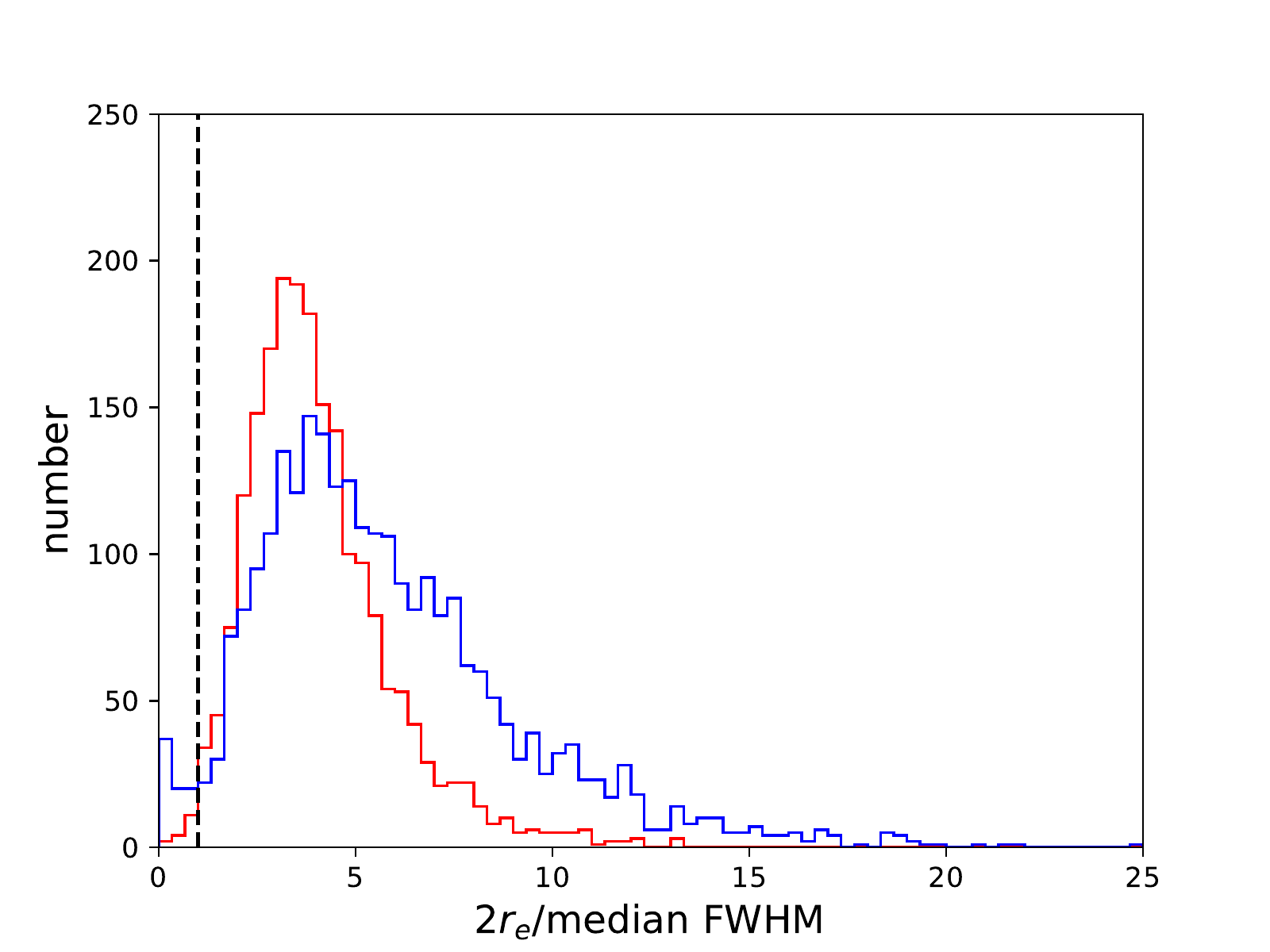}\hspace{-1.8em}
\includegraphics[width=.35\textwidth]{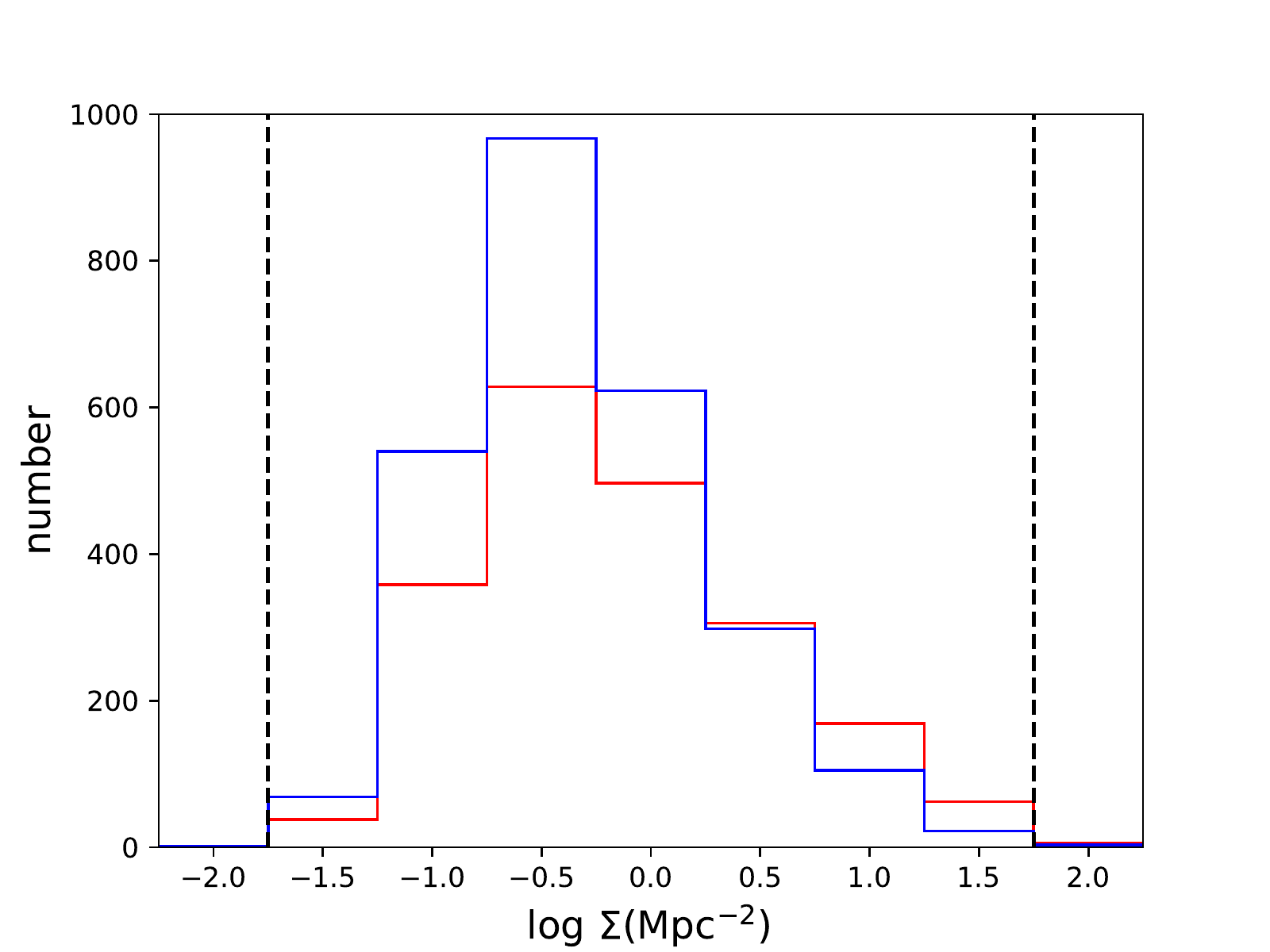}

\caption{Plots describing the selection criteria to obtain a final sample of disc galaxies.\textbf{Left:} The stellar mass distribution of parent sample. The vertical line at stellar mass $10^8 M_{\odot}$ defines the first selection cut on parent sample. Only galaxies more massive than $10^8 M_{\odot}$ are included in final sample. \textbf{Middle:} Distribution of ratio of bulge size (twice the half light radius $r_e$) and median PSF FWHM for galaxies of parent sample. The vertical line marks the position where this ratio is 1. Only those galaxies were chosen in final sample where bulge size is greater than size of the PSF. \textbf{Right:} Environmental distribution of galaxies in our parent sample. The galaxies falling inside the region bounded by lines at log $\Sigma$=$-1.75$ and log $\Sigma$=$1.75$ are included in our final sample.}

\label{fig:fig1}
\end{figure*}

Keeping in mind the structural similarity of classical bulges (CBs) to ellipticals and pseudobulges(PBs) to discs, one can use the two bulge types to divide disc galaxies into two structural classes. The first category will be of secularly evolved PB + disc system which will be a system closer to pure discs in their structure. And the second category of galaxies having a spheroidal CB + disc will form a structural class intermediate between disc dominated galaxies and pure spheroidal elliptical galaxies. Since both S0s and spirals are known to host both types of bulges, in additional to visual morphology one can use disc galaxy structural classes as defined above as measures of structural morphology. It is expected that the use of structural morphology will give new insights on visual Hubble morphological classes. For example, our previous study \citep{Mishra2018} on bulges of disc galaxies in a fixed environment has shown that the classical bulge hosting spirals are more likely to transform into S0 galaxies as compared to the pseudobulge hosting spiral galaxies. Therefore, one can hope that, recasting the observed trend of disc galaxy morphology with local environmental density in terms of structural morphological classes might help us to better understand the formation and the assembly of S0 galaxies in different environments.\par

In this work, we have attempted to understand, from an observational point of view, the formation and the distribution of S0 galaxies in different environments. Starting with a sample of spirals and S0s spanning a large range of environment, we have studied their environmental distribution by making use of structural morphology and their star formation properties. The organisation of this paper is as follows, Section \ref{sec:data} describes our data and sample selection. In Section \ref{sec:results} we present our results which are discussed in Section \ref{sec:dis} before we present the summary and conclusions in Section \ref{sec:sum}. Throughout this work, we have used the WMAP9 cosmological parameters: $H_0$=69.3 km s$^{-1}$ Mpc$^{-1}$, $\Omega_{m}$ = 0.287 , $\Omega_{\Lambda}$ = 0.713.

\begin{table*}

\centering
	\caption{Table indicating the galaxies in our final sample and their physical parameters which are used for our study in the manuscript. The columns shown in the table are (from left to right) SDSS Name, morphological T type, bulge surface brightness, bulge half light radius, environmental density parameter, galaxy half light radius, galaxy stellar mass, dark matter halo mass, $NUV-r$ colour and the type of the bulge. The units are shown below name of the each column. The blank spaces denote unavailable measurement of that particular quantity. A full version of this table is available in the electronic version of this paper.  }
	\label{tab:0}
	
	\begin{tabular}{|l|r|r|r|r|r|r|r|r|r|}
\hline
  \multicolumn{1}{|c|}{SDSS Name} &
  
  \multicolumn{1}{c|}{T} &
  \multicolumn{1}{c|}{$\mu_b(<r_e)$} &
  \multicolumn{1}{c|}{$r_e$} &
  \multicolumn{1}{c|}{log($\Sigma$)} &
  \multicolumn{1}{c|}{$R_{gal}$} &
  \multicolumn{1}{c|}{log($M_{*}$)} &
  \multicolumn{1}{c|}{log($M_{halo}$)} &
  \multicolumn{1}{c|}{$NUV-r$} &
  \multicolumn{1}{c|}{Bulge type} \\%%%%%%%%%%%%%%%
  
  \multicolumn{1}{|c|}{} &
  \multicolumn{1}{c|}{} &
  \multicolumn{1}{c|}{mag/arcsec$^2$} &
  \multicolumn{1}{c|}{kpc} &
  \multicolumn{1}{c|}{(Mpc$^{-2}$)} &
  \multicolumn{1}{c|}{kpc} &
  \multicolumn{1}{c|}{(M$_{\odot}$)} &
  \multicolumn{1}{c|}{(M$_{\odot}$)} &
  \multicolumn{1}{c|}{} &
  \multicolumn{1}{c|}{} \\

  \hline
  J155341.74-003422.84  & 3 & 20.082 & 2.66 & -0.062 & 9.32 & 11.083 & 12.949 & 2.822 & classical\\
  J154514.39+004619.89  & 5 & 23.038 & 3.17 & -0.89 & 4.71 & 9.173 & ----- & 0.958 & pseudo\\
  J112408.63-010927.83  & 0 & 18.341 & 1.19 & 0.0 & 2.99 & 10.645 & 12.127 & 2.868 & classical\\
  J113057.91-010851.06  & -2 & 18.705 & 1.57 & 0.0 & 3.57 & 10.507 & 12.27 & ----- & classical\\
  J120155.64-010409.34  & 1 & 19.655 & 2.9 & 0.0 & 19.38 & 10.606 & 11.722 & 3.549 & classical\\
  J131236.98-011151.02  & 3 & 20.701 & 4.76 & 0.0 & 12.06 & 11.057 & 14.02 & 2.569 & classical\\
  J140722.05-010546.97  & 0 & 20.458 & 3.12 & 0.268 & 7.24 & 10.865 & 13.838 & 4.305 & classical\\
  J110127.21-004305.93  & -2 & 18.804 & 2.81 & 0.0 & 5.36 & 11.082 & 12.604 & ----- & classical\\
  J110431.98-004349.84  & -3 & 17.853 & 0.56 & 0.0 & 1.39 & 9.985 & ----- & ----- & classical\\
  J124632.10-003836.77  & 0 & 19.461 & 5.62 & 0.324 & 10.64 & 11.301 & 13.601 & 4.786 & classical\\
\hline\end{tabular}

\end{table*}

\section{Data and sample selection}

\label{sec:data}

In order to carry out our study, we wanted to construct a statistically meaningful sample of galaxies with available information on morphology, structure and environment. We did this by making use of two catalogues containing large number of galaxies drawn from the SDSS. The first catalogue is by \cite{Nair2010}, which provides the visual morphological classification for 14034 spectroscopically targeted galaxies in the SDSS DR4 \citep{McCarthy2006}. We obtained structural information for each galaxy by cross matching with the second catalogue -- \cite{Simard2011} -- which contains bulge+disc decompositions in the SDSS $g$ and $r$ bands for a sample of 1,123,718 galaxies from the SDSS DR 7 \citep{Abazajian2009}. The cross match between these two catalogues resulted in 12,063 galaxies. Out of these 12,063 galaxies, we selected the disc galaxies having reliable bulge + disc fits. The detailed information on our selection criteria of reliable bulge + disc fits and the catalogues used is given in our previous works \citep{Mishra2017b, Mishra2018}, interested readers are requested to look there for details. We have also chosen to discard the galaxies which host a bar, as \cite{Simard2011} do not fit a bar component in their decompositions. Inclusion of such galaxies can lead to errors in estimation of the true bulge properties. Application of these cuts on original sample of 12,063 galaxies leaves us with a sample of 4692 objects. We refer to these 4692 galaxies as the parent sample. \par

In order to obtain the final sample of disc galaxies, we have applied three selection cuts on our parent sample. The first selection cut is applied on the stellar mass distribution of the galaxies where we have chosen to retain only those galaxies which have stellar mass $> 10^{8}$ $M_{\odot}$. The estimates of stellar mass for our sample  are taken from \cite{Kauffmann2003}. By making use of measured absorption line indices -- $D_n(4000)$ and $H \delta_{A}$ --, and broadband photometry they have developed a method of deriving the maximum-likelihood estimates of the stellar masses of about $\sim10^5$ SDSS galaxies. The application of first selection cut on our sample was done to remove the low mass dwarf galaxies which are not the objects of interests in our study. The second selection cut comes from the sizes of the bulges of disc galaxies in our parent sample. Previous works (\cite{Gadotti2009} and references therein) with galaxy image decompositions have shown that if the angular size of the bulge is smaller than the PSF size, one cannot get reliable estimates of bulge properties. Keeping this in mind, we have chosen to retain only those galaxies in our parent sample which have their bulge size (quantified by 2$r_e$, where $r_e$ is bulge half light radius) greater than the median PSF of 1.43 arcsec of the SDSS imaging \citep{Abazajian2009}. The mass and the bulge size distribution of the parent sample along with the selection cuts are displayed in the left and the middle panel of Fig. \ref{fig:fig1}.\par

The third selection cut comes from the need to have a statistically significant number of Spiral and S0 galaxies in different bins of the environment. We have obtained the measurement of the environment from \cite{Baldry2006}. Using the distance to the N'th nearest neighbour, they calculate the local environmental density around a galaxy. This process is subjected to a constraint that the neighbours must be found within a redshift range of $\Delta zc = \pm$1000 km/s centred around the redshift of that particular galaxy. This environmental density parameter is defined as $\Sigma_N$= N/$\pi d_{N}^{2}$, where $d_N$ is the distance to the N'th nearest neighbour. \cite{Baldry2006} provides the best estimate environmental density ($\Sigma$) by averaging the environmental densities $\Sigma_4$ and $\Sigma_5$. This estimate is known to vary only a little when its defining parameters are varied and so, we have chosen this parameter to quantify environment in our work. The environmental distribution of spiral and S0 galaxies of our parent sample along with the selection cut on the environment is shown in the right panel of Fig. \ref{fig:fig1}. We have chosen to restrict our sample within the environmental range of $-1.75\leq$ $log(\Sigma)$ $\leq1.75$ as this is the regime where we have statistically significant and comparable number of Spiral and S0 galaxies in each density bin. The application of all three selection criteria on the parent sample leaves us with a final sample of 4573 galaxies, out of which 2541 are spirals and rest 2032 are S0 galaxies. Throughout this paper, the term disc galaxies will be used to collectively refer to the combined population of spiral and S0 galaxies. The final sample of disc galaxies along with the estimates of their stellar masses, density parameters, bulge parameters and other quantities relevant for our study are listed in Table~\ref{tab:0}.\par

\section{Results}

\label{sec:results}
\subsection{Bulge classification and statistics}

The bulges in our final sample of galaxies were identified using their position on the Kormendy diagram. The Kormendy diagram is a plot of average surface brightness of the bulge within its half light radius ($\langle\mu_b (< r_e)\rangle$) against the logarithm of bulge half light radius ($r_e$). Elliptical galaxies are known to obey a tight linear relation on the Kormendy diagram. Classical bulges being structurally similar to ellipticals are expected to follow the same scaling relations as elliptical galaxies. Pseudobulges, being structurally different lie away from the scaling relation defined by ellipticals. A criterion to identify pseudobulges using previously mentioned arguments was given by \cite{Gadotti2009}. According to this criterion, all bulges which deviate more that three times the r.m.s. scatter from the best fit relation for elliptical galaxies are classified as pseudobulges while the bulges falling within this scatter are classified as classical bulges. This physically motivated criterion has also been used in recent works \citep{Vaghmare2013,Zahid2017,Mishra2017b}. \par

We have made use of the Kormendy relation from our previous work\citep{Mishra2017b} where the relation was obtained by fitting SDSS $r$ band photometric data of elliptical galaxies. The equation for best fit line is \\

$\langle\mu_b (< r_e)\rangle$ = $(2.330 \pm 0.047)$ log($r_e$) + $(18.160 \pm 0.024)$
\\

\begin{figure}
\includegraphics[width=\columnwidth]{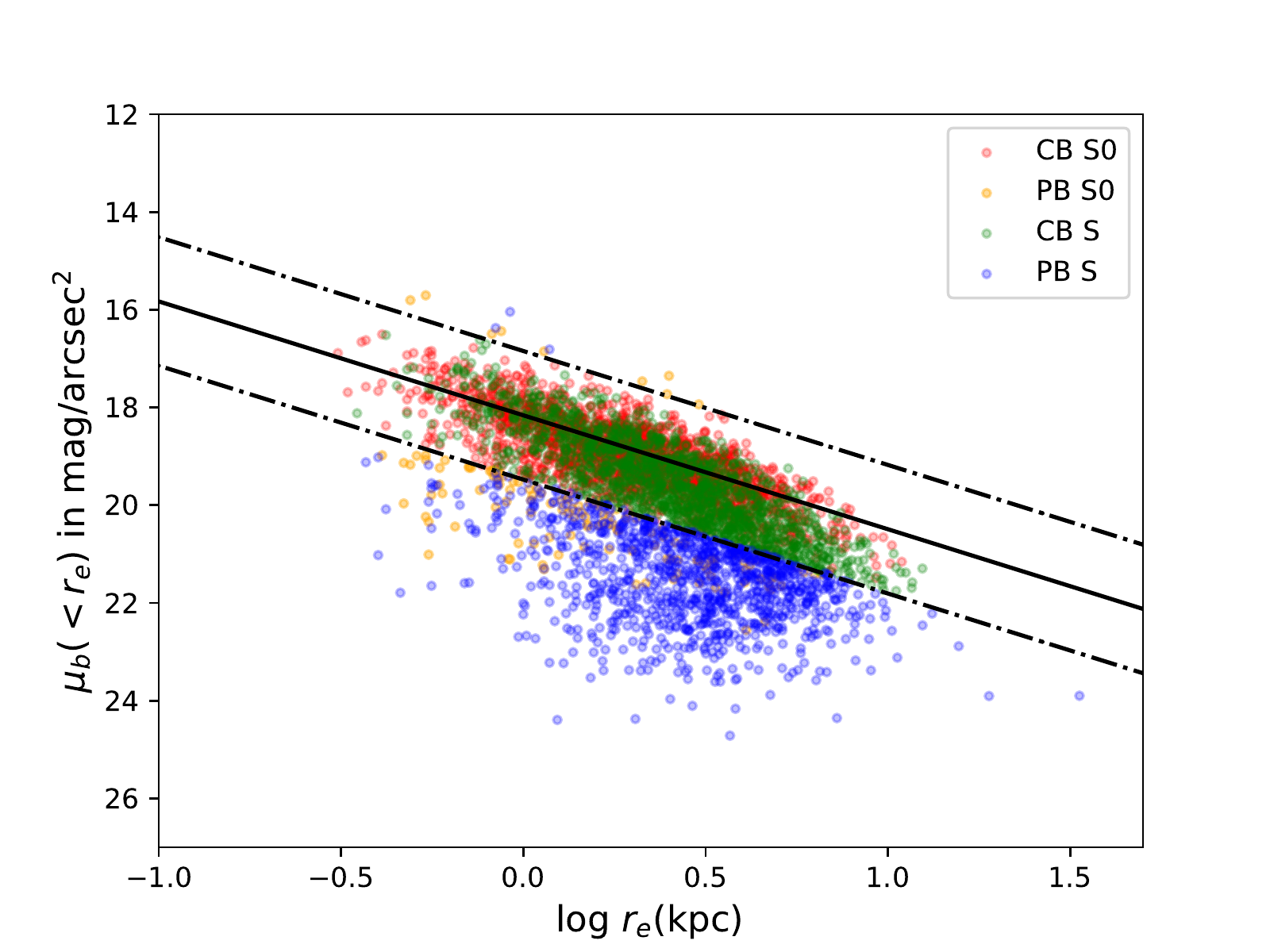}
\caption{Position of bulges of spiral and S0 galaxies on the Kormendy diagram. The solid line is the best fit line to the ellipticals and two dashed lines mark the boundary of 3$\sigma$ scatter as taken from \protect\cite{Mishra2017b}. The red and green points are the classical bulges hosted by S0 and spiral galaxies respectively. The pseudobulges of S0 and spirals are denoted by yellow and blue points respectively.}
\label{fig:Kormendy}
\end{figure}

The rms scatter in $\langle\mu_b (< r_e)\rangle$ around the best fit line is 0.429. All galaxies which lie more than 3 sigma scatter away from this line are classified as pseudobulge hosts while those within this scatter are classified as classical bulge hosts. For the S$\'e$rsic light profile, one can use the information on the magnitude and radial scale to infer average surface brightness as shown in \cite{Graham2005}. We have used this relationship to calculate the average surface brightness of the bulge. After carrying out classification of bulges hosted by spirals and S0 galaxies in our final sample, we find that out of 2541 spiral galaxies, 1526 (60.0\%) are classical bulge hosts while the rest 1015 (40.0\%) spirals are pseudobulge hosts. In case of total 2032 S0 galaxies, the number of classical and psuedobulge hosts are 1894 (93.2\%) and 139 (6.8\%) respectively. The position of classical and pseudobulges on the Kormendy digram for our final sample is shown in Fig. \ref{fig:Kormendy}. The type of bulge hosted by each one of the galaxies in our final sample is listed in Table~\ref{tab:0}.

\begin{figure*}
\includegraphics[width=\columnwidth]{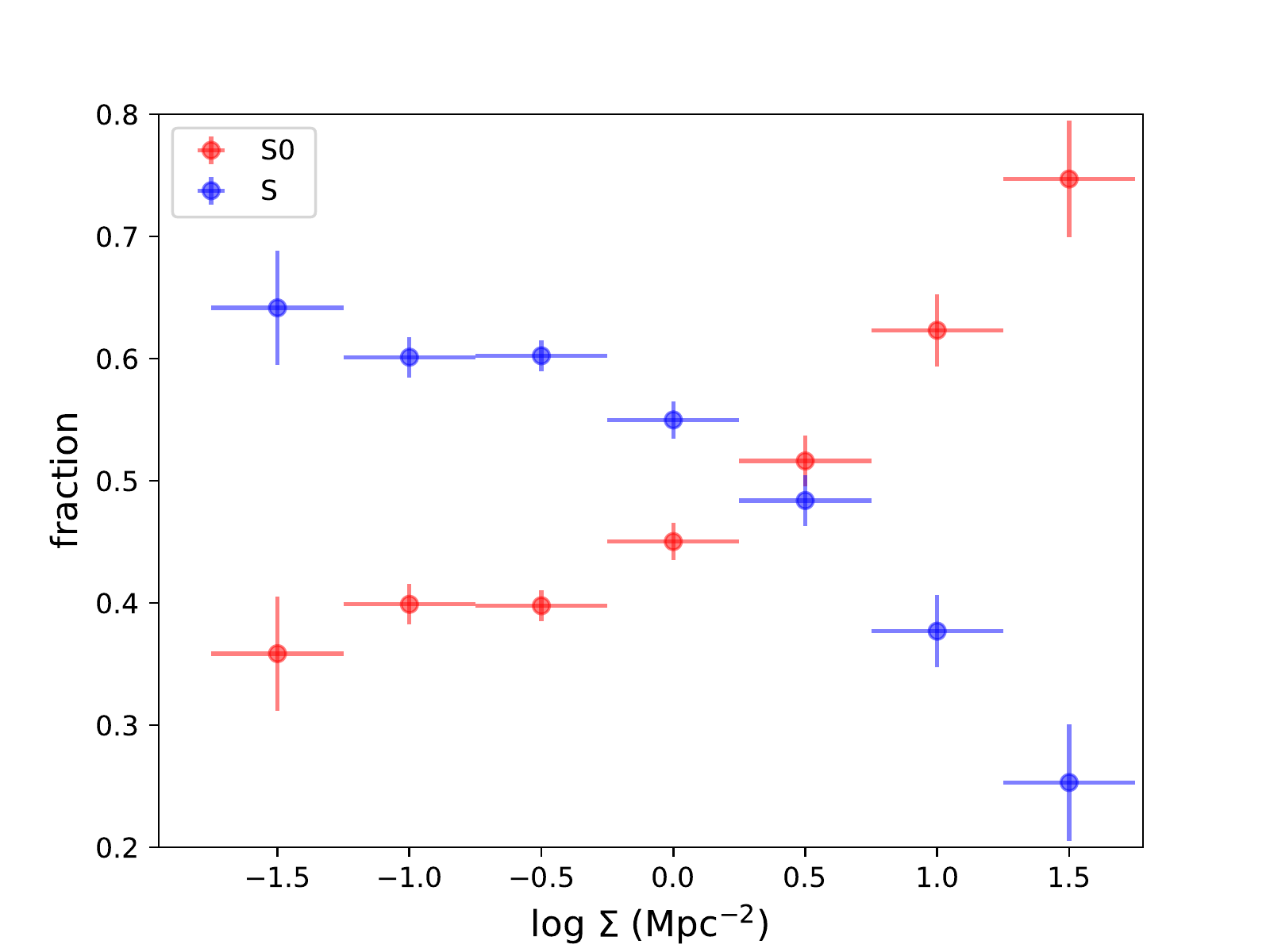}
\includegraphics[width=\columnwidth]{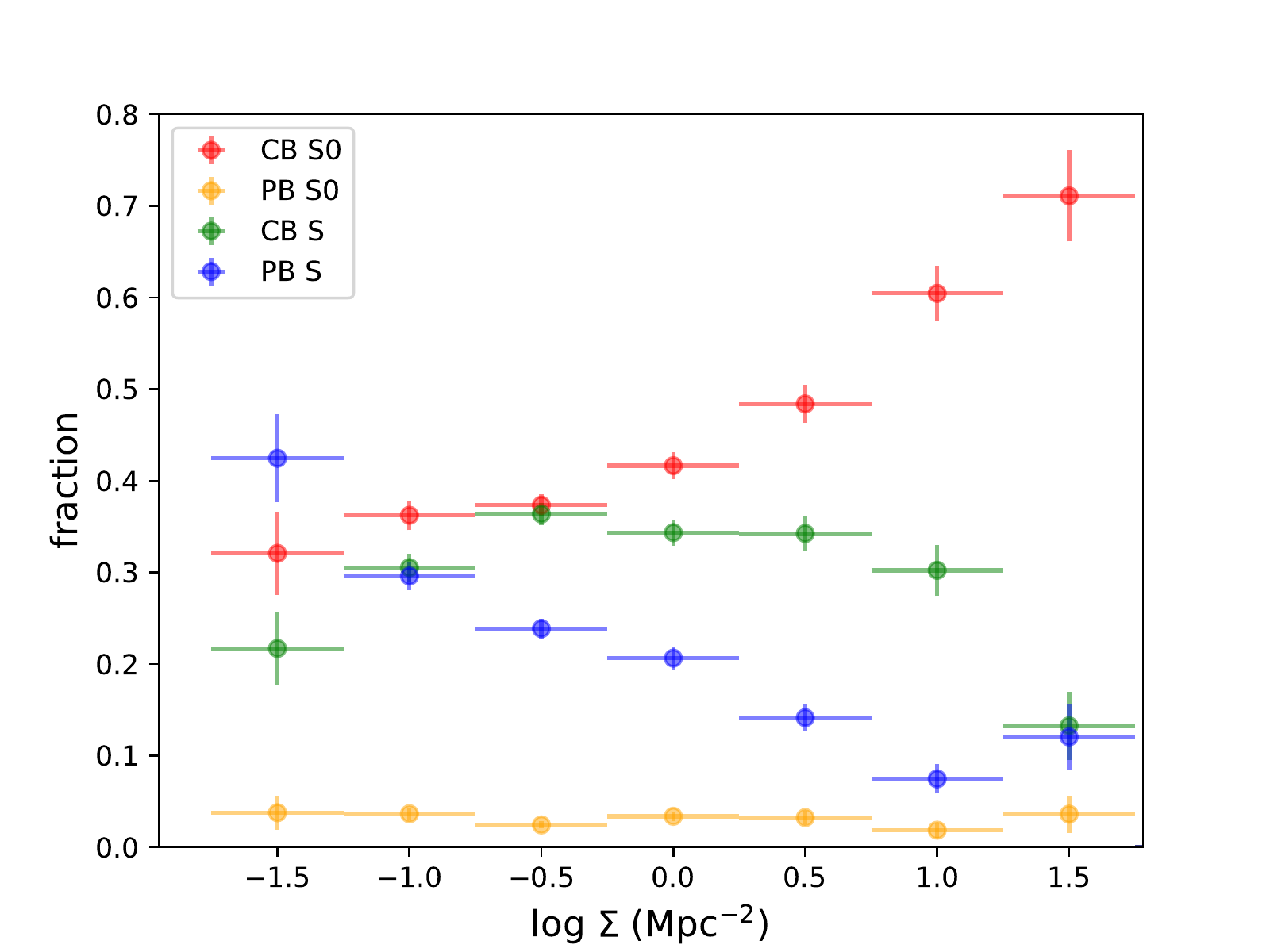}
\caption{Plots showing the environmental distribution of disc galaxy
  morphologies. The error bars on each point are Poisson
  errors. \textbf{Left:} plot of the fraction of spiral and S0
  galaxies as a function of environmental density. The red points are
  for S0 galaxies and the blue points are for spiral galaxies. In this
  plot, we note the known trend where one finds more S0 galaxies in
  high density environments. \textbf{Right:} fraction of classical and
  pseudobulge hosting spiral and S0 galaxies as a function of
  environmental density. The red, yellow, green and blue points refer
  to classical bulge hosting S0, pseudobulge hosting S0, classical
  bulge hosting spirals and pseudobulge hosting spirals
  respectively.  } 
\label{fig:fig2}
\end{figure*}

\subsection{Environmental distribution of disc galaxy morphologies.}

In order to understand the formation and the distribution of S0 galaxies in different environments, we begin our study by inspecting the morphological distribution of galaxies as a function of environment. The left panel of Fig. \ref{fig:fig2} shows the fraction of galaxies having spiral and S0 morphology in different bins of local environmental density. The figure shows that in low environmental density regime, majority of disc galaxies are spiral galaxies and the fraction of S0 galaxies is low. But as one goes towards higher density environments, the fraction of S0 galaxies increases monotonically such that at the highest density regime, most of the disc galaxies are S0 galaxies. The common occurence of S0 galaxies in high density environment is a well established result \citep{Dressler1980, Postman1984}, and is generally interpreted as being caused by higher conversion efficiency of spirals into S0 morphology in denser environments \citep{Smith2005, Park2007, Houghton2015}.\par

However, the question of morphological transformation becomes more interesting when one looks separately at the distribution of classical and pseudobulges in spiral and S0 galaxies. Looking at the statistics provided in the previous section, one can notice that the two classes of bulges are more or less equally common in spiral galaxies but most (93.2\%) of the S0 galaxies are classical bulge hosts. This fact is interesting in two ways. First, since S0 galaxies are thought to be transformed spirals, it is curious as to why there is a significant mismatch in fraction of bulge types seen in these two morphological classes. Second, since most of the S0 galaxies are classical bulge hosts, understanding their formation maximally contributes to the understanding of formation of S0 morphological class in general. In our previous study of bulges of disc galaxies in a fixed environment, we have shown that such a high classical bulge fraction seen in S0s can be explained by a preferential conversion of classical bulge hosting spirals into S0 galaxies \citep{Mishra2018}.\par

\begin{figure*}

%\centering
\includegraphics[width=.35\textwidth]{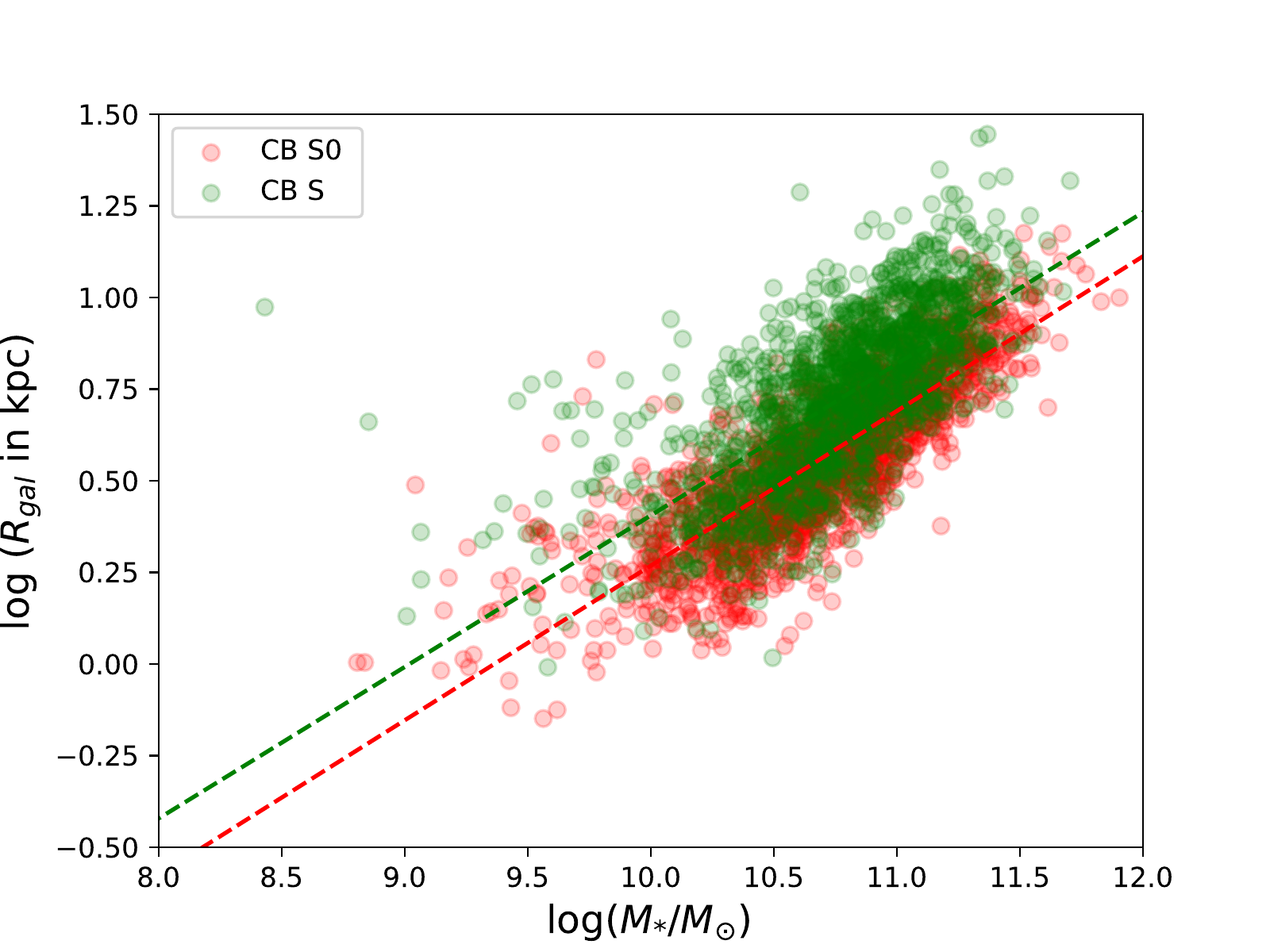}\hspace{-1.8em}
\includegraphics[width=.35\textwidth]{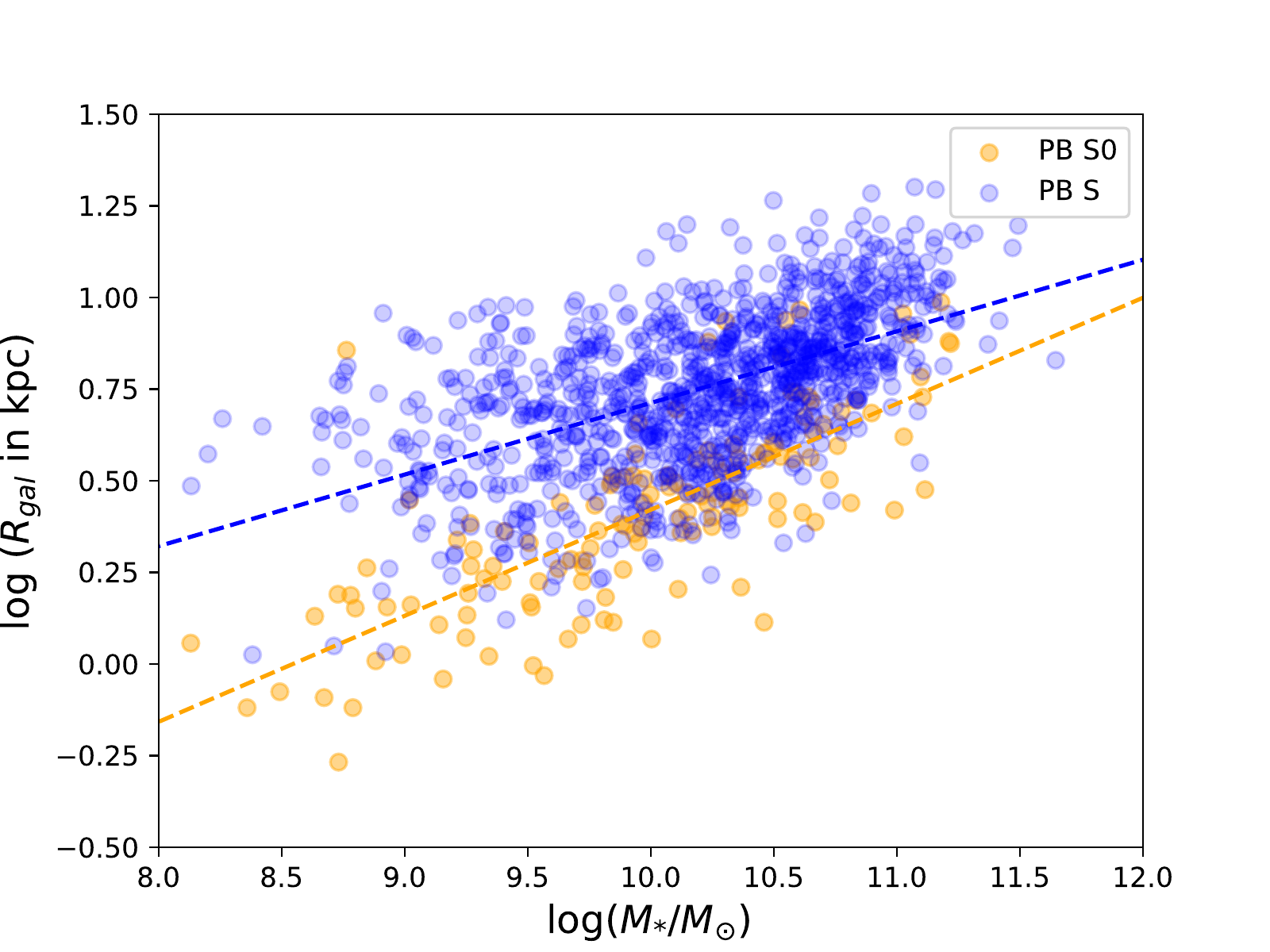}\hspace{-1.8em}
\includegraphics[width=.35\textwidth]{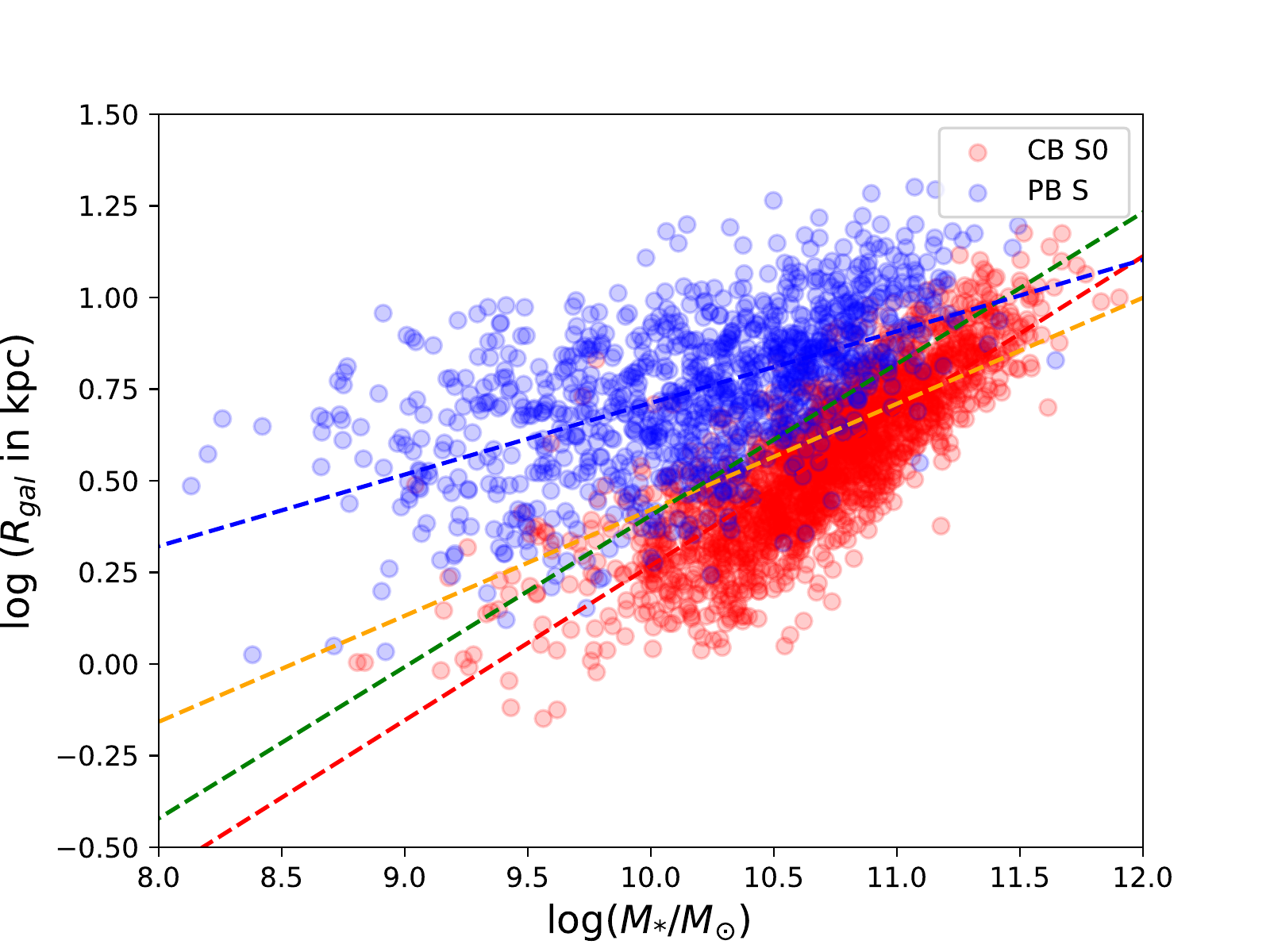}

\caption{The size-stellar mass relation for the classical and pseudobulge hosting disc galaxies in our final sample. The galaxy half light radius $R_{gal}$ is taken as a measure of galaxy size . The red, yellow, green and blue points refer to classical bulge hosting S0, pseudobulge hosting S0, classical bulge hosting spirals and pseudobulge hosting spirals respectively. The coloured lines are the best fit lines to the population of object denoted by each colour. \textbf{Left:} Size-stellar mass relation for classical bulge hosting spirals and S0 galaxies. The slopes of best fit lines for classical bulge hosting spirals and S0 are 0.41 and 0.42 respectively. The classical bulge hosting spiral and S0 galaxies have similar size-stellar mass relation with large overlap in parameters. This indicates that they form a structurally homogeneous class of objects with a possible common origin. \textbf{Middle:} Size-stellar mass relation for pseudobulge hosting spirals and S0 galaxies. For the pseudobulge hosting spirals and S0s the slope of the best fit line is 0.19 and 0.29 respectively. \textbf{Right:} Comparison of size-stellar mass relation of pseudobulge hosting spirals and classical bulge hosting S0 galaxies.}

\label{fig:fig3}

\end{figure*}

\begin{table*}
	\centering
	\caption{The slopes and the intercepts of best fit lines for classical and psudobulge hosting disc galaxies on size-stellar mass plane }
	\label{tab:1}
	%\resizebox{\columnwidth}{!}{%
	\begin{tabular}{lcccc} 
		\hline
		Galaxy class & Total number & Slope of best fit line & Intercept of best fit line & 1 $\sigma$ scatter around best fit line\\
		\hline
		
		Classical bulge hosting S0& 1894 & $(0.422 \pm 0.007)$& -3.951& 0.126\\
		Classical bulge hosting Spiral& 1526 & $(0.413 \pm 0.011)$& -3.728& 0.164\\
	    Pseudobulge hosting S0& 138 & $(0.289 \pm 0.021)$& -2.471& 0.169\\
		Pseudobulge hosting Spiral& 1015 & $(0.195 \pm 0.009)$& -1.241& 0.173\\
		\hline

	\end{tabular}
	%}
\end{table*}

Motivated by the above result, we wanted to see if this preferential conversion is valid in all environments. If it is valid, then how does it connect with the trend where S0 fraction is a monotonically increasing function of environment. To explore along these lines, we have reformulated the disc galaxy morphology-density relation by complimenting the visual morphology with structural information. More specifically, we further divide the spirals and S0 galaxies into those hosting classical and pseudobulge and plot their fraction as function of environmental density in the right panel of Fig.\ref{fig:fig2}. From this figure, one can see that the fraction of pseudobulge hosting spirals decrease monotonically with increase in density while the environmental distribution of pseudobulge hosting S0s is quiet flat. The most interesting trend in this plot is, however, of the classical bulge hosting spirals and S0 galaxies. When viewed together in this plot, it occurs that both of these classes follow somewhat similar trend with environment in low density regime with an increase in fraction of both classes with environmental density. Interestingly, as one keeps going towards higher densities, the fraction of classical bulge hosting spirals takes a downward turn around density bin of log$\Sigma (Mpc^{-2})$=[-0.25,0.25]. In the same bin, we notice that the fraction of classical bulge hosting S0 galaxies goes on increasing with environmental density. The environment corresponding to log$\Sigma (Mpc^{-2})$=[-0.25,0.25] is typical of galaxy groups. There are 1119 galaxies in our sample residing in this environmental range. We have found the mean and median richness of the group to which these galaxies belong, using the available group membership information from \cite{Yang2007}. The values for the mean and median group richness comes out to be 5 and 2 respectively. This increase in classical bulge hosting S0 galaxies seems to be occurring at the expense of classical bulge hosting spirals at higher densities. This statement is still speculative at this stage and indirectly assumes that only classical bulge hosting spirals are getting converted to S0 galaxies. One must carefully examine the structural and star formation properties of all the morphological classes in order to make a stronger connection between S0 galaxies and their possible progenitors. We begin by examining the structural properties of disc galaxies in the following section.

\subsection{Structural properties of classical and pseudobulge hosting disc galaxies}

We first compare the classical and pseudobulge host galaxies in terms of their global structure. We do so by examining their size-mass relation. Fig. \ref{fig:fig3} shows the size-mass relation for the classical bulge (CB) and pseudobulge (PB) hosting spirals and S0 galaxies in our sample. The galaxy half light radius (which we denote by $R_{gal}$) is taken as a measure of galaxy size from \citet{Simard2011}. The estimates of galaxy half light radius are provided in Table \ref{tab:0}. The mass on the x-axis is the stellar mass of the whole galaxy. A straight line has been fit to each population. The slopes, intercepts and the 1 $\sigma$ scatter around the best fit line  of size-mass relation for classical and pseudobulge hosting galaxies are given in Table \ref{tab:1}.\par

Inspection of Fig. \ref{fig:fig3} coupled with best fit size-mass relation for the galaxies in our sample is revealing. The size-mass relation for classical bulge hosting spirals and S0 galaxies is  plotted in the left panel of Fig. \ref{fig:fig3}. One can see from this plot that classical bulge hosting spirals and S0 follow a similar relation on the size-mass plane. The slope of best fit line for classical bulge hosting spirals and S0 galaxies are 0.413 and 0.422 respectively. We also find a large overlap in structural parameters of classical bulge hosting spiral and S0 galaxies. This tells us that classical bulge hosting spirals and S0 galaxies are structurally similar, differentiated just by visual appearance. Such similarity in the structural morphology points towards their possible common origin. The middle panel of Fig. \ref{fig:fig3} shows the size--mass relation for the pseudobulge hosting spiral and S0 galaxies. One can notice that pseudobulge hosting galaxies follow a different and less steep size-mass relation as compared to their classical bulge hosting counterparts. The slope of best fit line of size-mass relation for pseudobulge hosting spirals and S0 galaxies is 0.195 and 0.289 respectively. In the right panel of Fig. \ref{fig:fig3}, we have compared the structural properties of pseudobulge hosting spirals and classical bulge hosting S0 galaxies on the size-mass plane. The best fit line for size-mass relation for classical bulge hosting spirals and pseudobulge hosting S0 galaxies are also shown in the same plot. It is clear that pseudobulge hosting spirals are bigger in size than classical bulge hosting S0 galaxies at similar stellar masses. This result makes it unlikely that pseudobulge hosting spirals are the major contributors in giving rise to S0 population. We provide reason to support this claim in the following paragraph. \par 

\begin{figure*}

%\centering
\includegraphics[width=.355\textwidth]{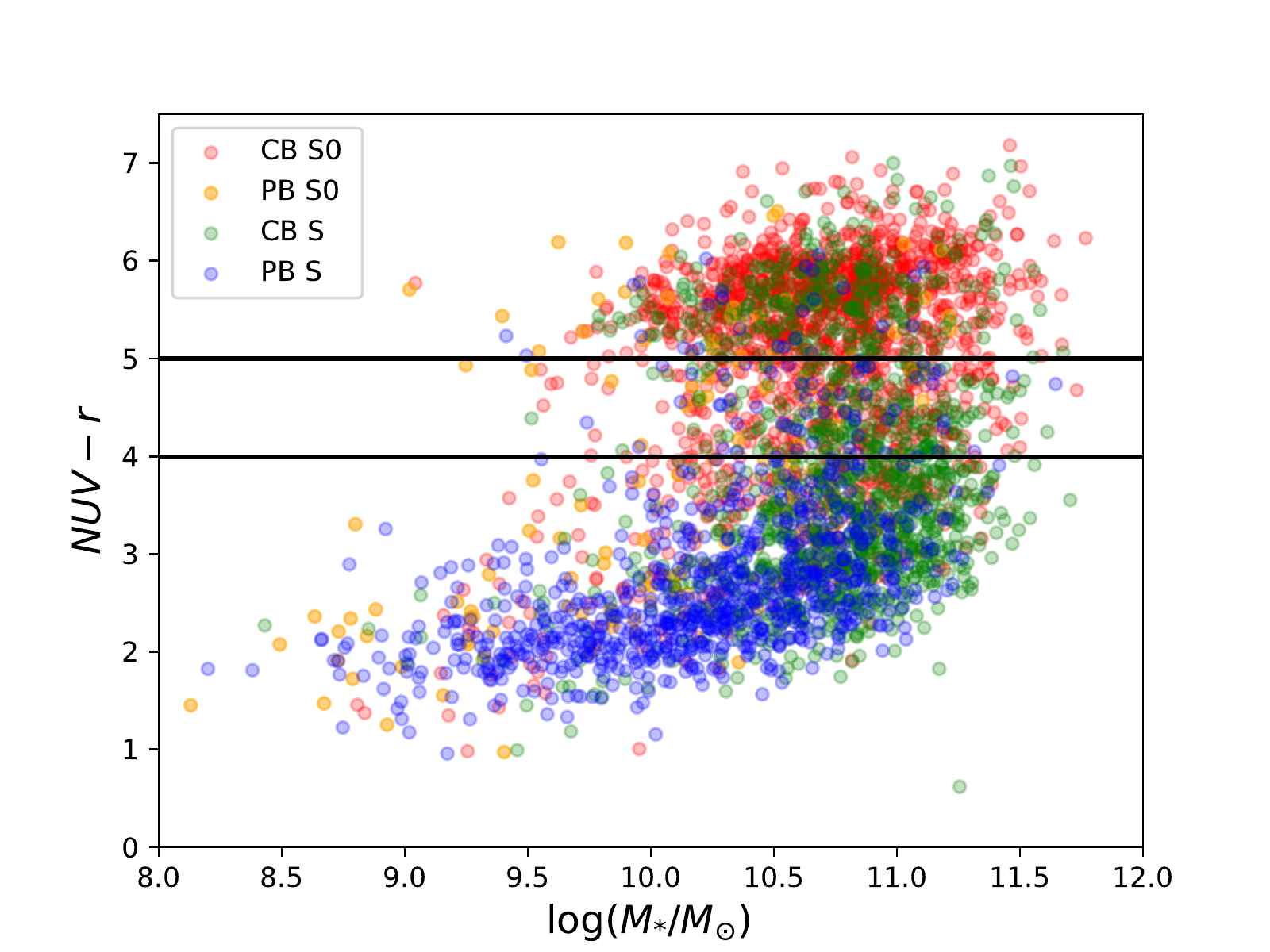}\hspace{-2.4em}
\includegraphics[width=.355\textwidth]{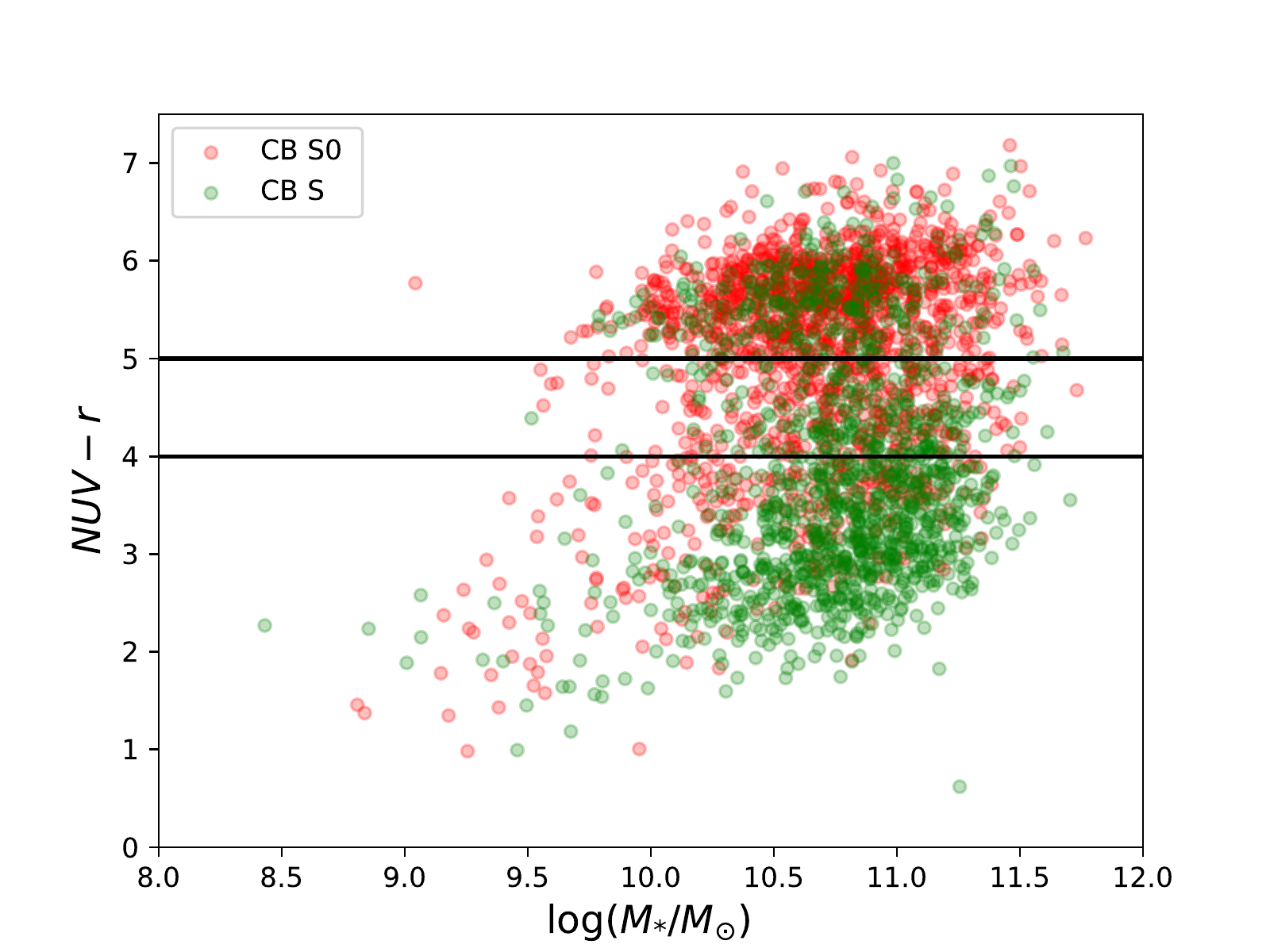}\hspace{-2.35em}
\includegraphics[width=.355\textwidth]{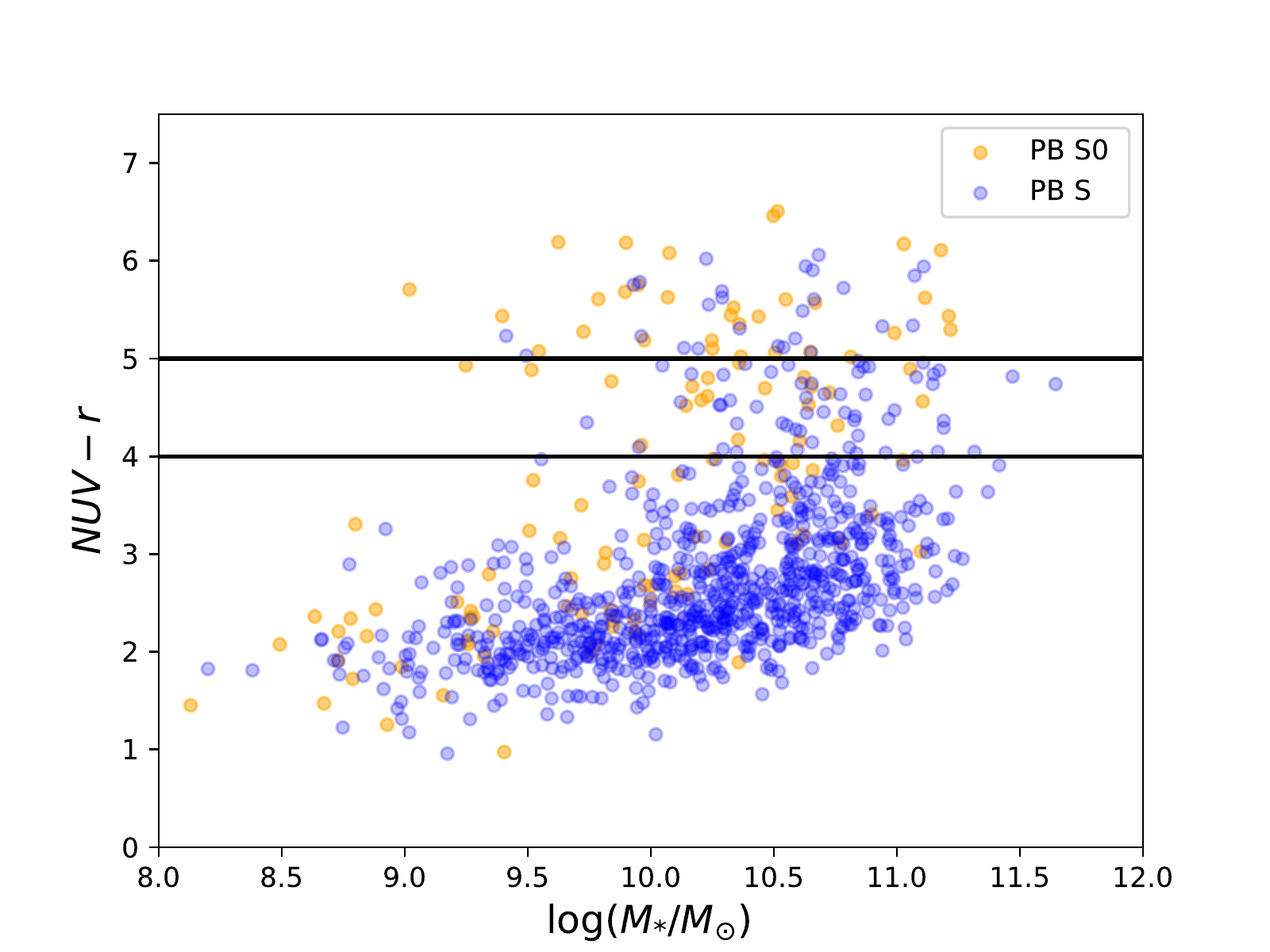}

\caption{The global star formation property of spirals and S0 galaxies
  as traced by NUV-r colour-stellar mass diagram.  The horizontal
  lines at $NUV = 5$ and $NUV = 4$ mark the boundary of the green
  valley region \citep{Salim2014} which separates the quenched red
  sequence lying above the green valley from the star forming galaxy
  sequence which lies below this region. The colour scheme is the same
  as in the right panel of Fig. \ref{fig:fig2}. The left panel shows
  the global star formation properties of all the disc galaxies taken
  together. For the clarity of presentation, the NUV-r colour-stellar
  mass plot for classical and pseudobulge hosting galaxies is
  separately shown in the middle and right panels respectively.}
\label{fig:fig4}

\end{figure*}

\begin{table*}
	\centering
	\caption{A summary of global starformation properties of population of classical and pseudobulge hosting disc galaxies. Only the galaxies having available $NUV-r$ colour measurements (3790 out of 4573) in the final sample are presented here. We have divided population of each galaxy classes into categories of quenched ($NUV-r$ $\geq 5$), green valley ($4<NUV-r<5$) and starforming ($NUV-r$ $\leq 4$) galaxies. }
	\label{tab:2}
	%\resizebox{\columnwidth}{!}{%
	\begin{tabular}{lcccc} 
		\hline
		Galaxy class & Total number & Quenched (\%)& Green Valley (\%)& Starforming (\%) \\
		\hline
		
		Classical bulge hosting S0& 1561 & 961 (61.9\%) & 318 (20.4\%) & 276 (17.7\%)\\
		Classical bulge hosting Spiral& 1264 & 288 (22.8\%) & 221 (17.5\%) & 755 (59.7\%)\\
	    Pseudobulge hosting S0& 120 & 32 (26.6\%) & 20 (16.7\%) & 68 (56.7\%)\\
		Pseudobulge hosting Spiral& 845 & 26 (3.1\%) & 54 (6.4\%) & 765 (90.5\%)\\
		\hline

	\end{tabular}
	%}
\end{table*}

 A spiral galaxy can be transformed to an S0 galaxies though galaxy mergers \citep{Tapia2017}. Galaxy mergers are also known to give rise to classical bulges\citep{Brooks2015}, therefore one can argue that mergers can transform pseudobulge hosting spirals into classical bulge hosting S0 galaxies. However, galaxy mergers always end up increasing the size of the remnant galaxy and hence if majority of S0s are forming this way they must have sizes either similar or greater than the sizes of pseudobulge hosting spiral population which is not the case as shown in Fig. \ref{fig:fig3}. But mergers are not the only way to form S0 galaxies and hence one must also take into account other possible paths of S0 formation before ruling out the possibility of conversion of pseudobulge hosting spirals into S0 galaxies. For example, it has been suggested in the literature \citep{Furlong2017} that tidal stripping can reduce the size of satellite galaxies. Therefore in principle, satellite spiral galaxies can get converted into smaller S0 galaxies via tidal stripping followed by quenching. However, the majority (about 78\%) of galaxies in our final sample are central (defined as most massive galaxy within a group from \citet{Yang2007}) galaxies where the argument for tidal stripping driven conversion may not be applicable. Furthermore, we also have checked the size-mass relation of only satellite galaxies in our sample, and found it to be consistent with result of  Fig. \ref{fig:fig3}. We find that the pseudobulge hosting satellite spirals are still larger in size compared to the satellite S0 galaxies. If tidal stripping was transforming satellite pseudobulge hosting spirals into S0 galaxies, they would have had similar size-mass relation but we do not see such behaviour. Therefore, it seems unlikely that majority of S0 galaxies have formed out of pseudobulge hosting spirals. On the other hand, we have found that the populations of classical bulge hosting spirals and S0 galaxies show a large overlap in the size-mass plane even when the sample is divided into different bins of environmental densities. A simple shutdown of starformation can transform the classical bulge hosting spirals into a population of S0 galaxies with similar size-mass relation. Therefore, it is likely that classical bulge hosting spirals are the main progenitors of population of S0 galaxies. This argument naturally motivates us to explore the star formation properties of classical and pseudobulge hosting disc galaxies. \par

Before moving further, we would like to comment on the curious case of pseudobulge hosting S0 galaxies. This class of galaxies have size mass relation intermediate between classical bulge hosting discs and pseudobulge hosting spirals. The intermediate behaviour shown by pseudobulge hosting S0 galaxies is interesting in its own right but due to the low number statistics it is difficult to draw conclusive inference about their peculiar behaviour on the size-mass plane. Adopting a viewpoint where we wish to understand the most significant contributor to the dominant population of S0 galaxies, namely the ones which host a classical bulge, we postpone the discussion on the pseudobulge hosting S0 galaxies towards the end of this paper.\par

\subsection{Star formation properties of classical and pseudobulge hosting disc galaxies}

\begin{figure*}
\includegraphics[width=0.355\textwidth]{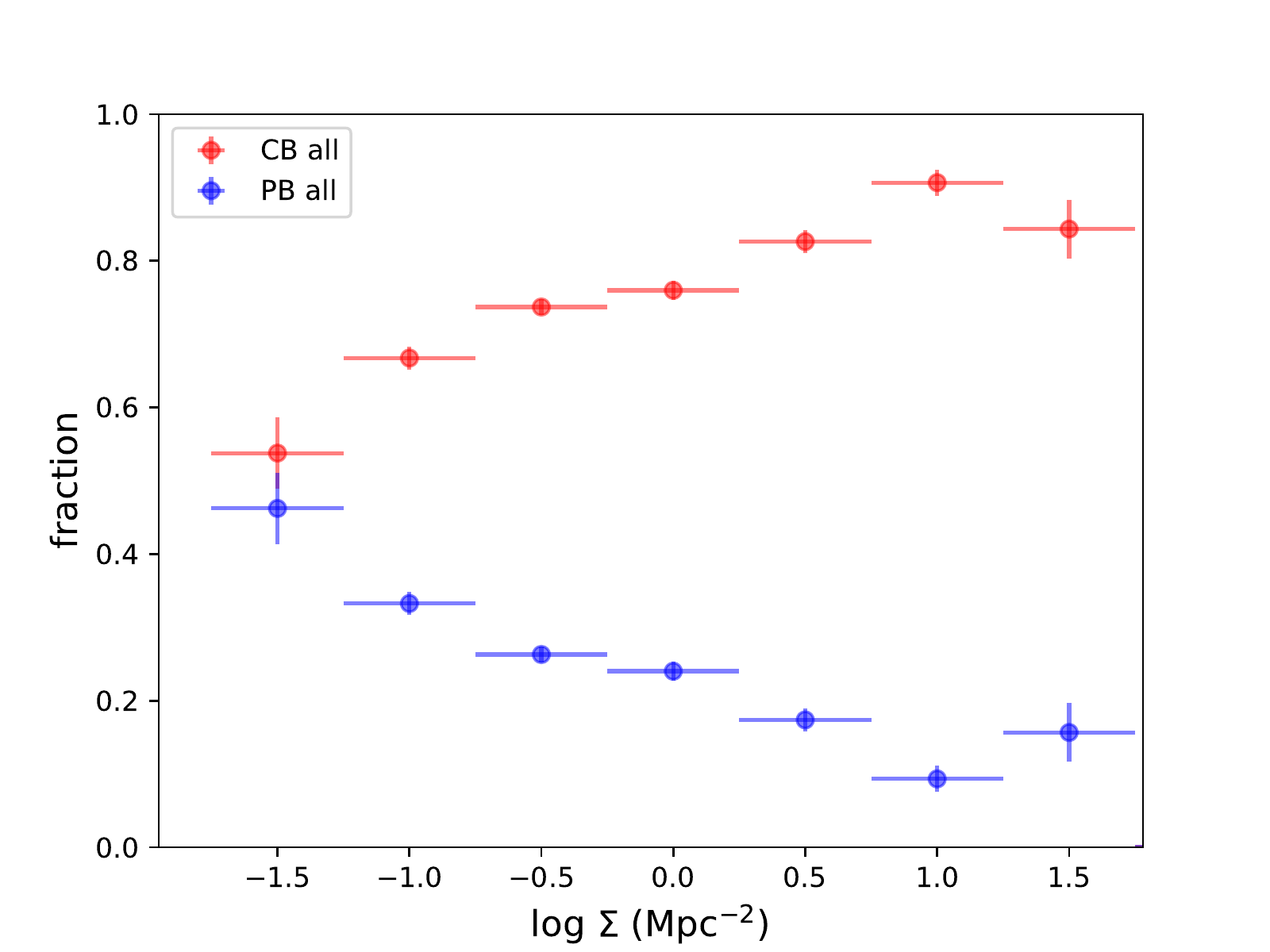}\hspace{-2.4em}
\includegraphics[width=0.355\textwidth]{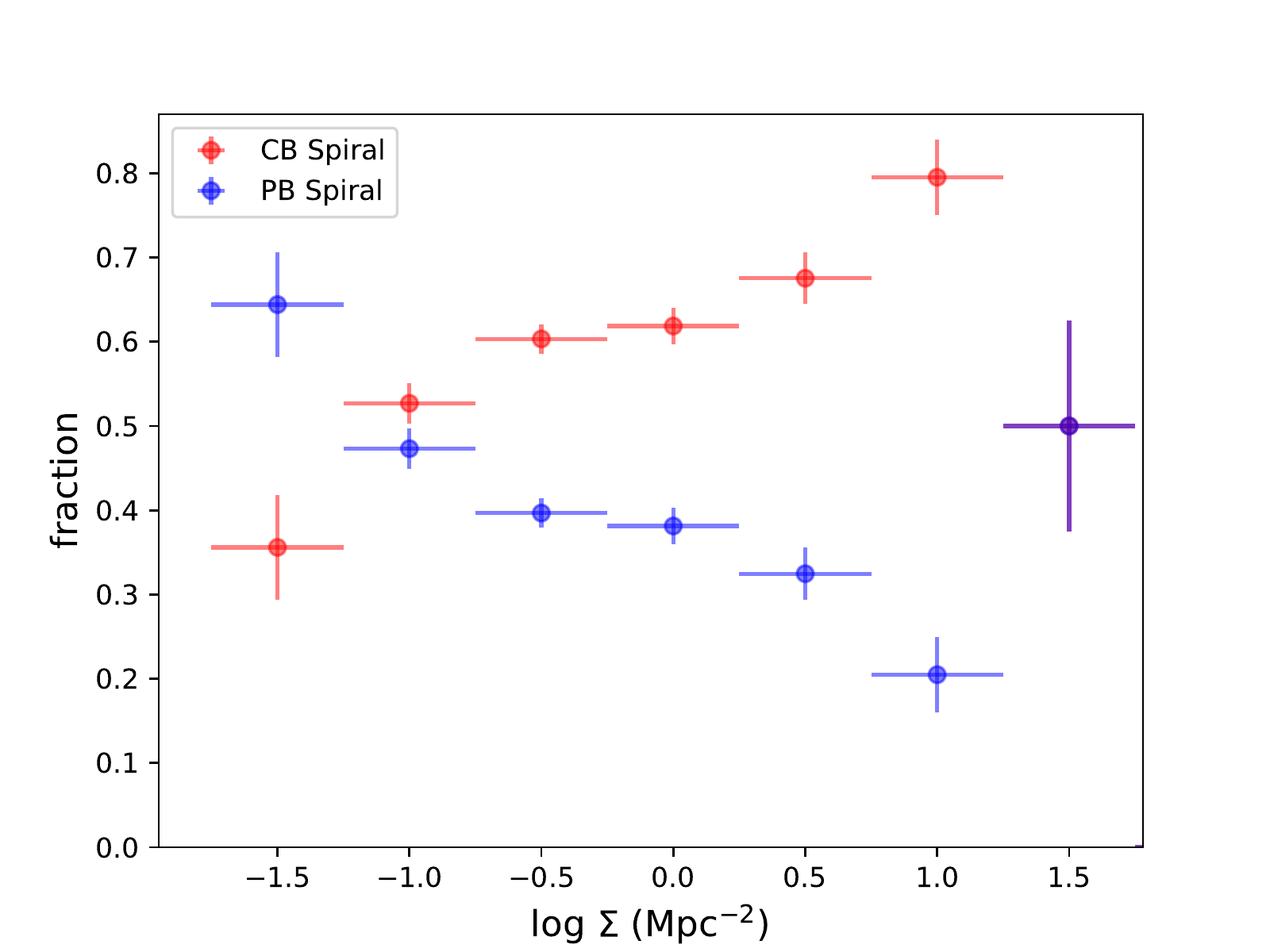}\hspace{-2.35em}
\includegraphics[width=0.355\textwidth]{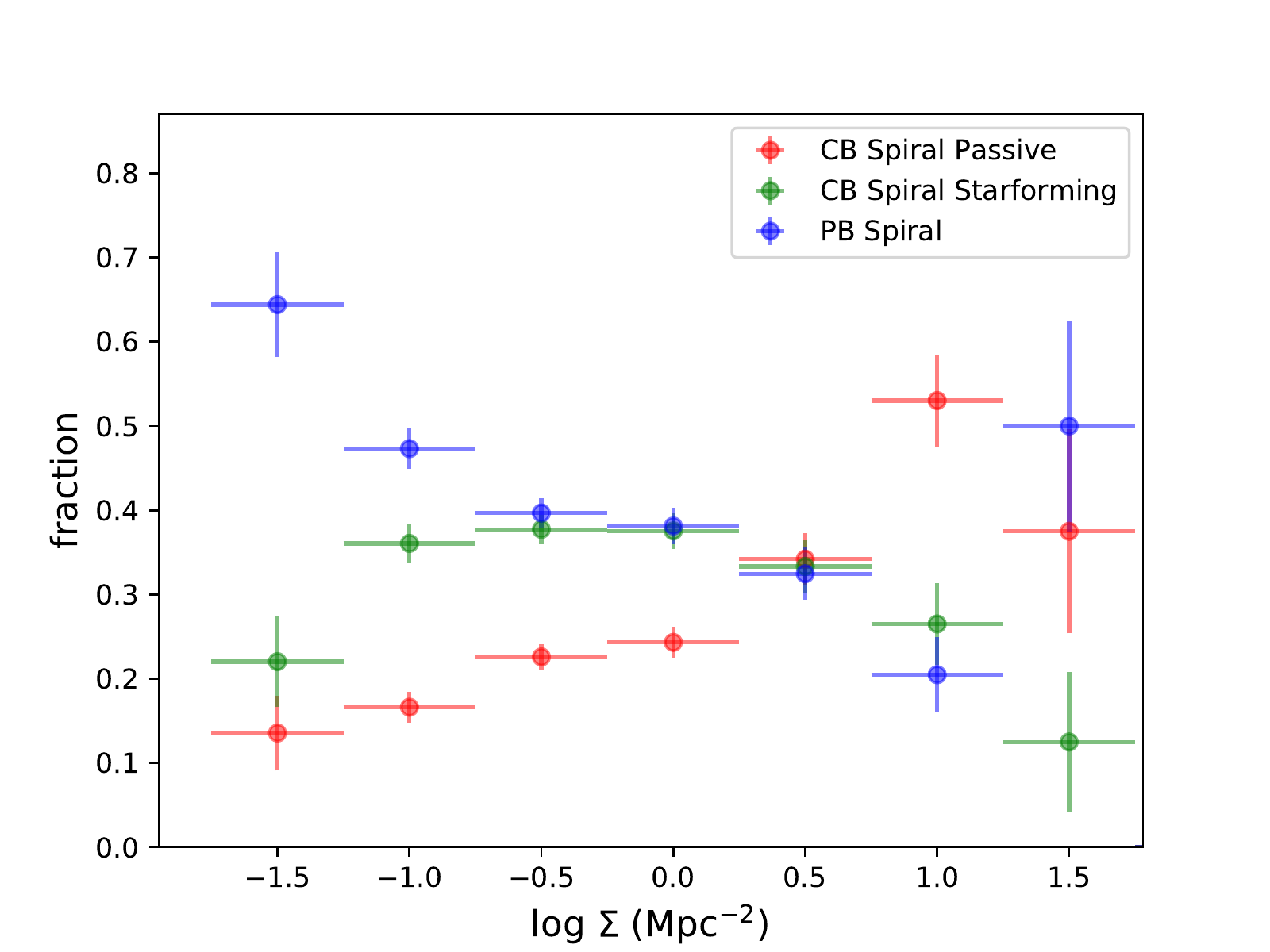}
\caption{Plots showing the dependence of disc galaxy structure on the environment. \textbf{Left:} The structural morphology density relation for disc galaxies irrespective of their visual morphology. The fraction of disc galaxies hosting a classical or a pseudobulge is plotted as function of environmental density. The red and the blue points are for classical and pseudobulge hosting disc galaxies respectively. \textbf{Middle:} Plot showing fraction of spiral galaxies hosting a classical or a pseudobulge as function of environmental density. The red and the blue points are for classical and pseudobulge hosting spiral galaxies respectively. \textbf{Right:} Similar plot as the middle panel but now we have divided classical bulge hosting spirals into passive ($NUV-r>4$) and star forming ($NUV-r <=4$) classes. The passive and the star forming spirals hosting a classical bulge are denoted by red and green points respectively.}
\label{fig:fig5}
\end{figure*}

To gain further insight on the formation of S0 galaxies, we have checked the star formation properties of galaxies in our sample by plotting them on $NUV-r$ colour-stellar mass diagram. The SDSS $r$ band and $GALEX$ $NUV$ magnitude for the galaxies in our final sample were obtained from  the RCSED catalogue \citep{Chilingarian2017}. The RCSED catalogue is a value added catalogue of Spectral Energy Distributions (SED) of 800,299 Galaxies in 11 ultraviolet, optical, and near-infrared bands. The catalogue provides us with SDSS $r$ band petrosian magnitude. The $GALEX$ $NUV$ magnitude is estimated using the NUV flux from a Kron-like elliptical aperture. Out of total 4573 galaxies in our final sample, we have measurements of $NUV-r$ colours for 3790 galaxies from this catalogue which are provided in Table \ref{tab:0}. \par 

The left panel of Fig. \ref{fig:fig4} shows $NUV-r$ colour-stellar mass diagram for all classical and pseudobulge hosting spirals and S0 galaxies in our sample having available $NUV-r$ colour. In this diagram, the two horizontal lines at $NUV = 5$ and $NUV = 4$ mark the boundary of the green valley region \citep{Salim2014}. The galaxies lying above the green valley in the $NUV-r$ color-mass diagram are the quenched galaxies while those lying below the green valley are star forming. Table \ref{tab:2} summarises the global starformation properties of classical and pseudobulge hosting disc galaxies in our final sample.\par 

In the middle and the right panel of Fig. \ref{fig:fig4}, we have separately plotted $NUV-r$ colour-stellar mass plots for classical and pseudobulge hosting disc galaxies respectively. For these plots, one can see that pseudobulge hosting spirals are almost always star forming ($\sim$90\%) and hence might not have gotten chance to convert into S0 class via disc fading and disappearance of spiral arms. However, a significant fraction ($\sim$ 82\%) of classical bulge host spirals are either situated in green valley region or in the quenched sequence and have  the potential to acquire S0 morphology via spiral arm fading. A simple quenching of star formation which does not alter the size of the galaxy dramatically, can naturally explain the similarity of size-mass relation seen in classical bulge hosting spirals and S0 galaxies. Therefore, it seems likely that the classical bulge hosting spirals are the major contributors in giving rise to the S0 population. It should be noted that in the past \cite{Bait2017}, exploring the connection between the visual morphology and starformation properties of galaxies, have argued that early-type (Sa-Sbc) spirals get transformed into S0 galaxies via quenching. The median morphology class of the classical bulge hosting spirals in our final sample is of early-type Sab spirals which is consistent with their result. \par

Investigation of disc galaxy properties from structural and starformation point of view has given us clues on what kind of spirals get converted into S0 galaxies. However, we are yet to understand the reason behind the frequent occurrence of S0 galaxies in high density environments. We take up this issue in the next section.

\subsection{Dependence of structural morphology and quenching on environment}

As was mentioned in the introduction, galaxies that we see today have followed many different evolutionary paths since their formation. The difference seen with respect to their visual appearance, structure and assembly is largely governed by the different initial conditions of their progenitors and different interaction with the environment in which they reside. This results in a relation between morphology and environmental density such that among the disc galaxy classes, one finds more S0 galaxies in high density environment as compared to spirals. At this point, it seems that the classical bulge hosting spirals have a major contribution in giving rise to the population of S0 galaxies. But it is still not clear why S0 galaxies are more common in denser environments. One can speculate that perhaps to begin with the environmental distribution of progenitor of S0 galaxies is biased towards high density environment. It is also possible that processes acting in high density environments efficiently quench these galaxies and give rise to observed distribution of S0 galaxies.\par

We first check the dependence of disc galaxy structure on environment by plotting the environmental distribution of classical and pseudobulge disc galaxies, irrespective of their visual morphology, in the left panel of Fig. \ref{fig:fig5}. One can see from this plot that there exists a relation between disc galaxy structure and the environment such that the classical bulge hosting disc galaxies become increasingly common in high density environments. This motivates us further to look for a similar trend in case of the spirals, the supposed progenitors of S0 galaxies. In the middle panel of Fig. \ref{fig:fig5}, we have plotted the fraction of classical and pseudobulge hosting spirals as function of environmental density. Anticipating the need to study the trend of quenching with the environment, we have plotted only the galaxies which have available $NUV-r$ colour measurements. From this figure, it is clear that there exists a biased distribution where the fraction of classical (pseudo-) bulge host galaxies increases (decreases) with increase in environmental density.\par

Since classical bulge hosting spiral galaxies are more likely to end up in the S0 morphological class, it is not surprising that one will find more S0s in high density environments as compared to low density ones. However, one must also be able to explain the observed environmental trend of disc galaxy morphology in which the S0 galaxies become increasingly common as one goes from low to high density environments. We attempted to understand this by splitting the classical bulge hosting spirals into starforming and passive population. A galaxy is defined to be passive if its not in actively star forming, i.e. it is either located in quenched or green valley sequence of $NUV-r$ colour-stellar mass plane. We define the galaxies having their $NUV-r$ colour in range of $NUV-r<=4$ and $NUV-r>4$  as starforming and passive galaxies respectively. We have plotted the environmental distribution of passive and starforming classical bulge host spirals in the right panel of Fig. \ref{fig:fig5}. We did not make similar separation among pseudobulge hosting spirals galaxies as most ($>\sim$90\%) of them are starforming. Additionally, they are unlikely to get converted into S0 galaxies as discussed in previous sections. From this figure, one can notice that both the quenched and the star forming fraction of classical bulge host galaxies show a steady initial rise as one moves from low to mid density environments. In this regime, there are more star forming classical bulge hosting spirals than their passive counterparts. Moving towards higher densities, we notice that the star forming fraction of classical bulge hosting spirals makes a turn over and decreases steadily. At the same time, the quenched fraction of classical bulge hosting spiral galaxies becomes dominant and keeps on increasing with environmental density. This suggest that high density environment is more efficient in quenching the classical bulge hosting spiral galaxies.\par

One can use this result to qualitatively understand the relative environmental distribution of spiral and S0 galaxies. The low density environment is devoid of spirals which can potentially acquire S0 morphology. One finds this environmental regime to be dominated by spirals which are either structurally different (pseudobulge hosting spirals) or are still starforming and hence not suitable for morphological transformation to S0 population. Therefore, one sees more spirals than S0 galaxies in low density environment. This trend gets reversed at high density where one finds more classical bulge hosting spirals, majority of which are quenched and hence can potentially acquire S0 morphology. We conclude that the frequent occurrence of S0 galaxies in dense environments is mainly due to the combination of two factors. There exists an environmental distribution of progenitors of S0 galaxies which is biased towards higher densities and the high density environment can efficiently transform them into S0 galaxies via quenching.

\begin{figure*}
\includegraphics[width=0.355\textwidth]{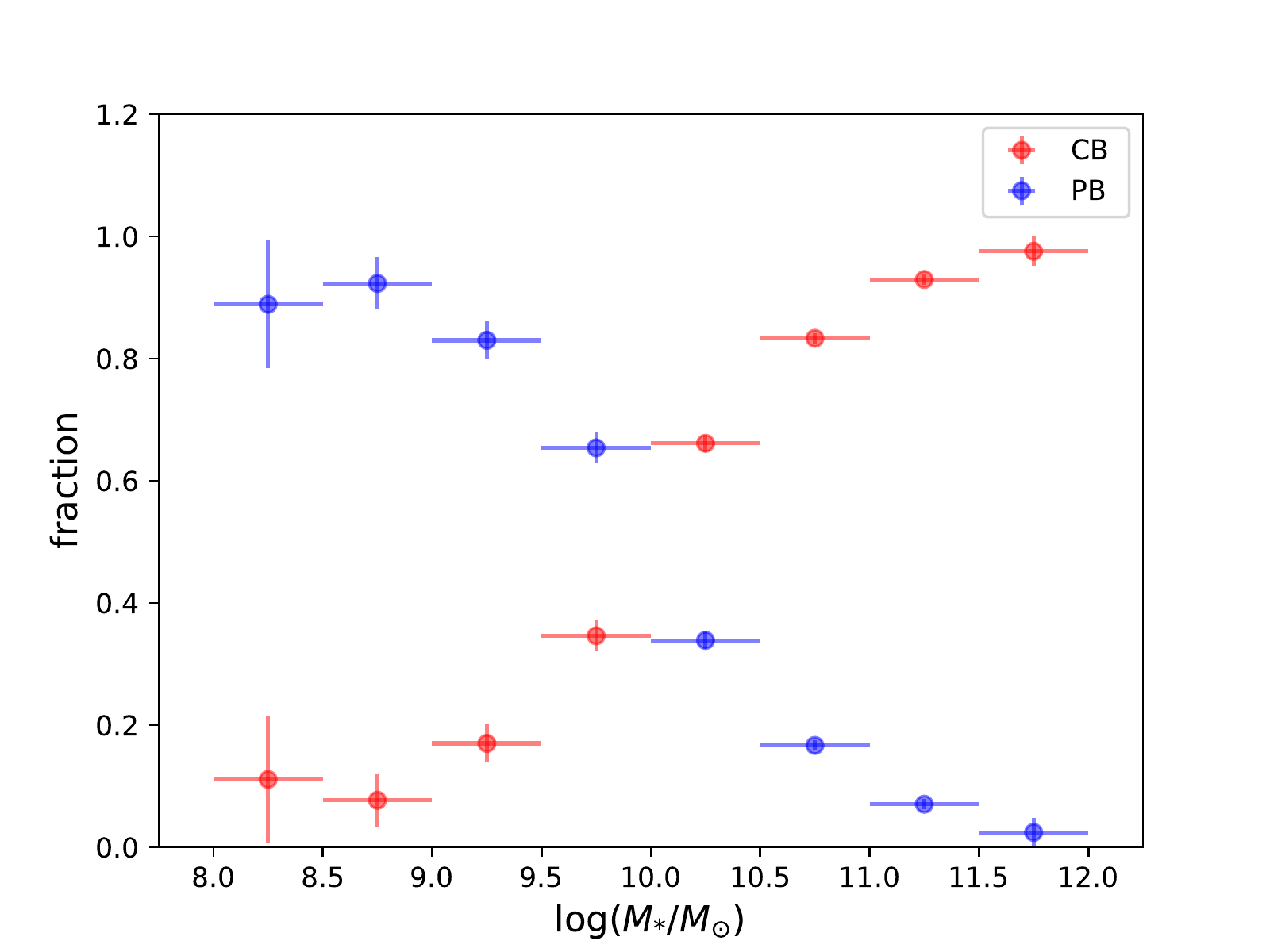}\hspace{-2.4em}
\includegraphics[width=0.355\textwidth]{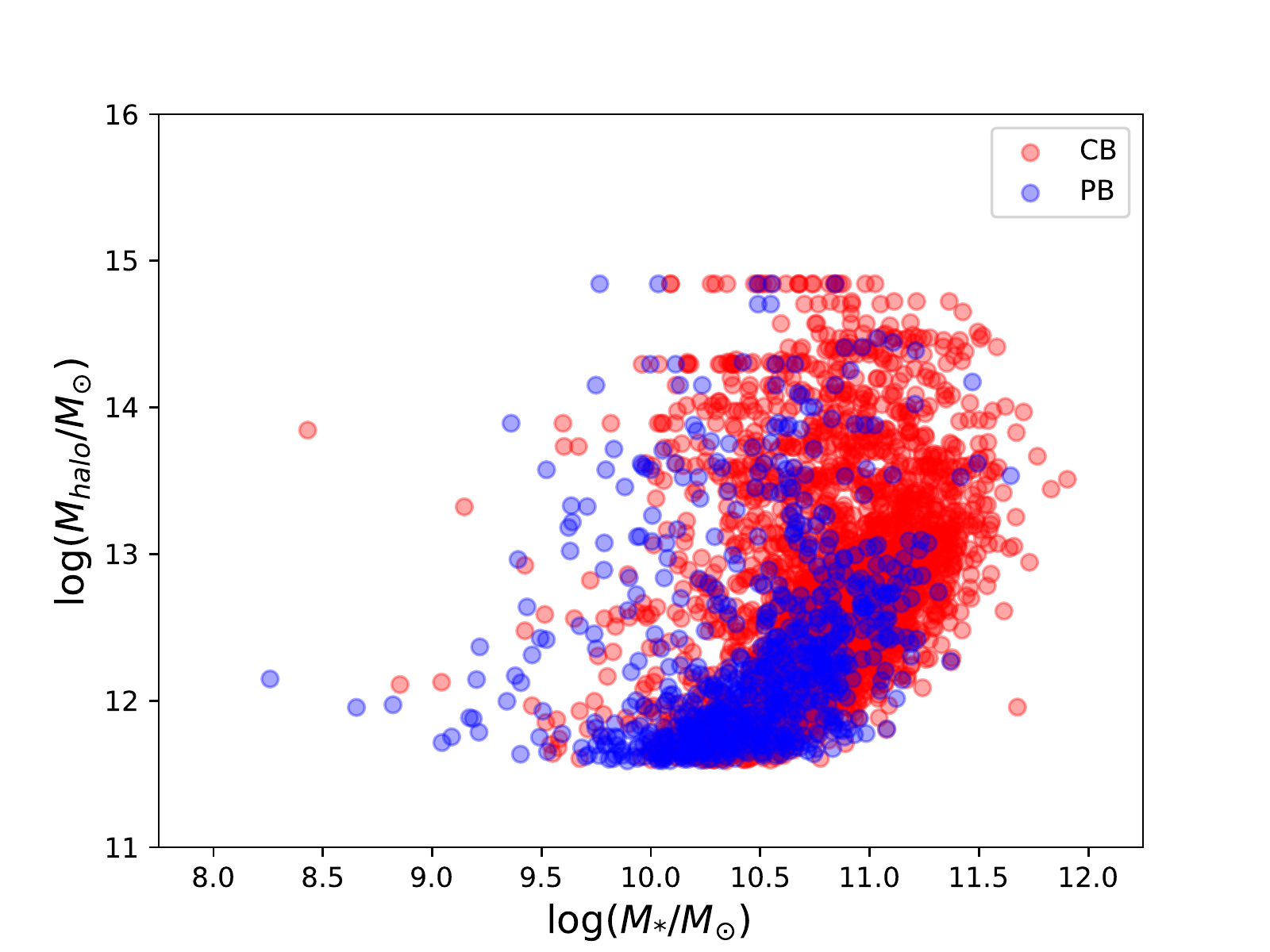}\hspace{-2.35em}
\includegraphics[width=0.355\textwidth]{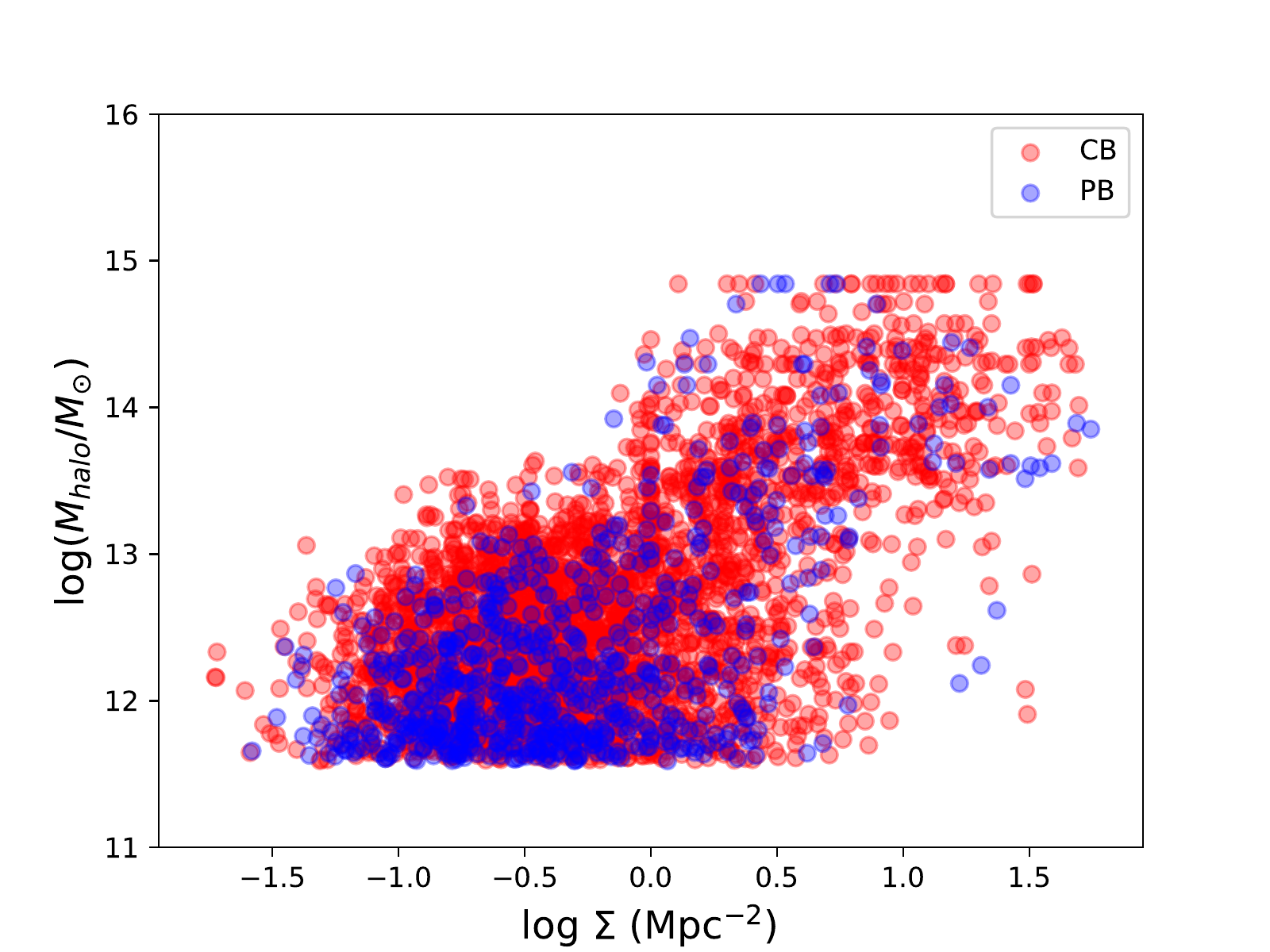}

\caption{Plots showing dependence of bulge fraction on host galaxy properties and those of its parent dark matter halo \textbf{Left:} Classical and pseudobulge fraction as a function of host galaxy stellar mass. The red points denote the classical bulge host galaxies while the blue points are for pseudobulge host galaxies. The plot shows that classical(pseudo-) bulges are more (less) common in massive galaxies. \textbf{Middle:} Scatter plot of classical and pseudobulge host galaxies on parent dark matter halo mass- galaxy stellar mass diagram. The color scheme is the same as in the left panel. This plot shows a mild correlation between the galaxy stellar mass and parent dark matter halo mass. \textbf{Right:} Scatter plot of parent dark matter halo mass and environmental density for the galaxies in our sample.}
\label{fig:fig6} 

\end{figure*}

\section{Discussion}

\label{sec:dis}

In the previous section, we inferred that the classical bulge hosting spirals are the main contributors in giving rise to the population of S0 galaxies. We also found that the observed dependence of S0 galaxies with environment is due to combination of two factors. The first factor is the presence of a relation between disc galaxy structure and the environment such that it gives rise to a biased environmental distribution of progenitors of S0 galaxies. The second factor is that the denser environment is more efficient in quenching the progenitor population and converting them to S0 galaxies. But we have not yet addressed the possible causes behind these two factors. We now take up these issues one by one.

\subsection{On the origin of disc galaxy structural morphology-density relation}

We first explore the possible reason for the existence of a relation between structural morphology and environment for disc galaxies as shown in the left panel of Fig. \ref{fig:fig5}. This plot tells us that, irrespective of their visual morphologies, there exists a relation purely between disc galaxy structure and the environment. The connection between disc galaxies with different bulge types and environment has also been reported by previous works \citep{Fisher2011, Kormendy2016, Wang2019} in the literature. We find that the classical bulge+disc systems become increasingly common as one goes towards higher environmental densities. The existence of a similar relation for spirals seems to be just a representation of this more fundamental relation between structural morphology and the environment. Understanding the reason behind the existence of such a relation between galaxy structure and environment requires answering a simple question: why do we find more classical bulge hosting disc galaxies in high density environment? However, first we should consider the fact that this relation may not be a direct one. It is possible that it might be just a result of correlation of structural morphology with some galaxy property which in turn is linked to the environment. From previous studies we know that classical bulges are more common in galaxies having higher stellar mass \citep{Fisher2011,Mishra2017a}. It is also known that galaxies having higher stellar masses tend to be hosted by high mass dark matter haloes \citep{Yang2012,Wechsler2018}. And since more massive haloes tend to reside in dense environments \citep{Haas2012,Lee2017}, it might be possible that one sees more classical bulge hosting disc galaxies in dense environments due to correlations between morphology, halo mass and the environment.\par

In order to check for this possibility, we have explored the possible dependence of bulge types on their host galaxy properties and its parent dark matter halo in Fig. \ref{fig:fig6}. In the left panel of Fig. \ref{fig:fig6}, we have plotted the classical and pseudobulge fraction as a function of galaxy stellar mass. It is clear from this plot that classical bulge fraction in disc galaxy population increases with increase in stellar mass, ie. one finds more classical bulge hosts among disc galaxies with higher stellar mass. The middle panel of Fig. \ref{fig:fig6} shows the scatter plot of stellar mass of classical and pseudobulge host galaxies vs. the mass of the dark matter halo in which each galaxy resides. One can see a mild correlation between stellar mass of galaxy and the halo mass. One can also notice majority of pseudobulge hosting galaxies reside in haloes having mass less that $10^{13} M_{\odot}$, while higher mass haloes are dominated by classical bulge hosts. At this point, we want to make it clear that the quoted mass of parent dark matter halo is not the dark matter content of each galaxy, but is the mass of dark matter halo which host the group in which that particular galaxy belongs. The mass estimates of parent dark matter halo of galaxies in our sample are taken from \citet{Yang2007} and are listed in Table \ref{tab:0}. They assign halo masses by first converting group luminosity to a mass by multiplying it with a certain mass to light ratio. The halo mass estimates are then iteratively refined using velocity dispersion measurements of member galaxies to measure dynamical halo mass. The right panel of Fig. \ref{fig:fig6} shows a significant correlation of the halo mass with environmental density. It is clear from this plot that the highest (log$\Sigma$>0.5) density regime of environment is dominated by haloes having mass greater than $10^{13} M_{\odot}$. Then it is possible, that correlation of bulge type with stellar or halo mass is making classical bulge hosting disc galaxies more common in denser environments.\par 

\begin{figure*}

%\centering
\includegraphics[width=.35\textwidth]{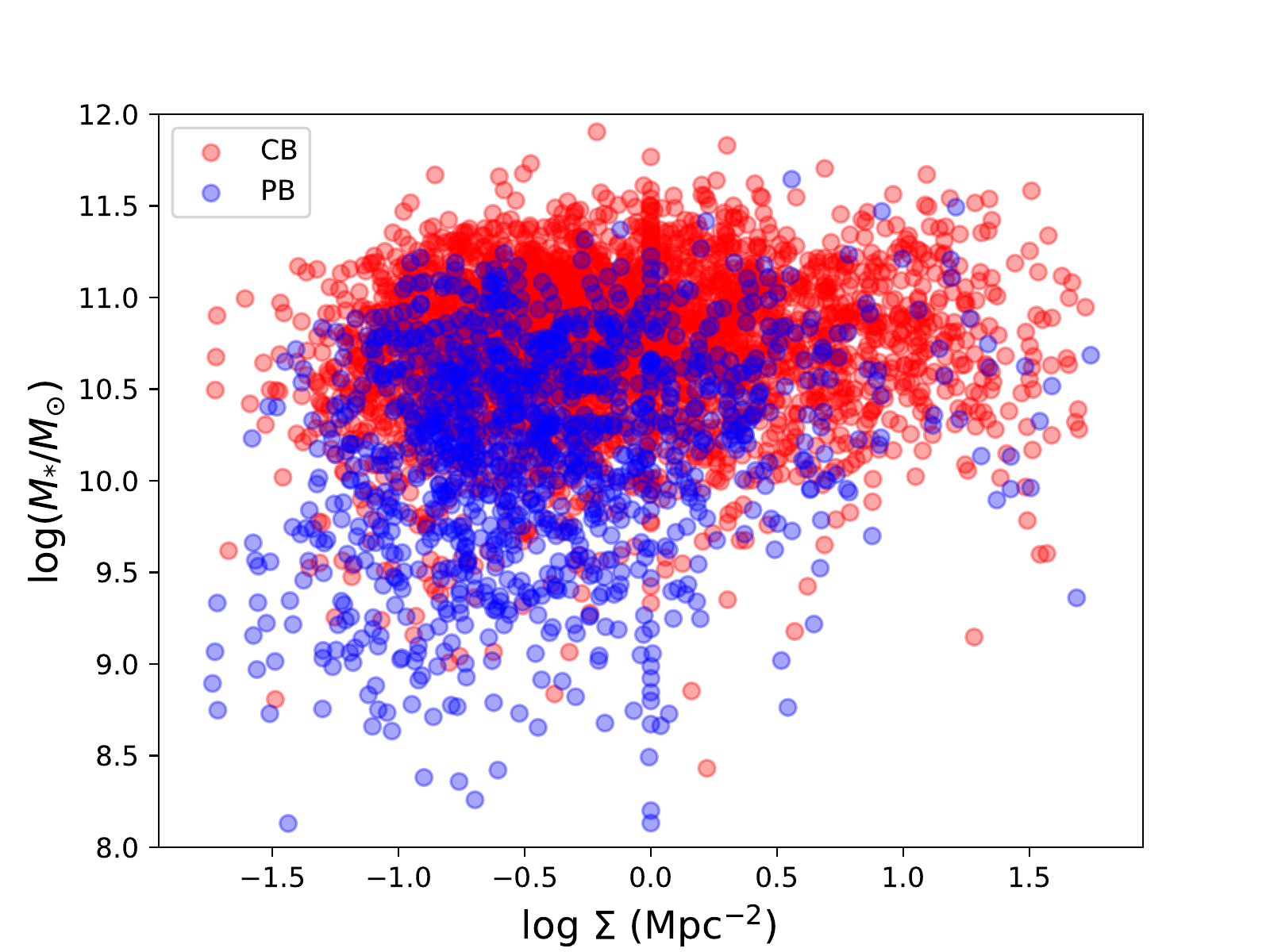} \hspace{-2.4em}
\includegraphics[width=.35\textwidth]{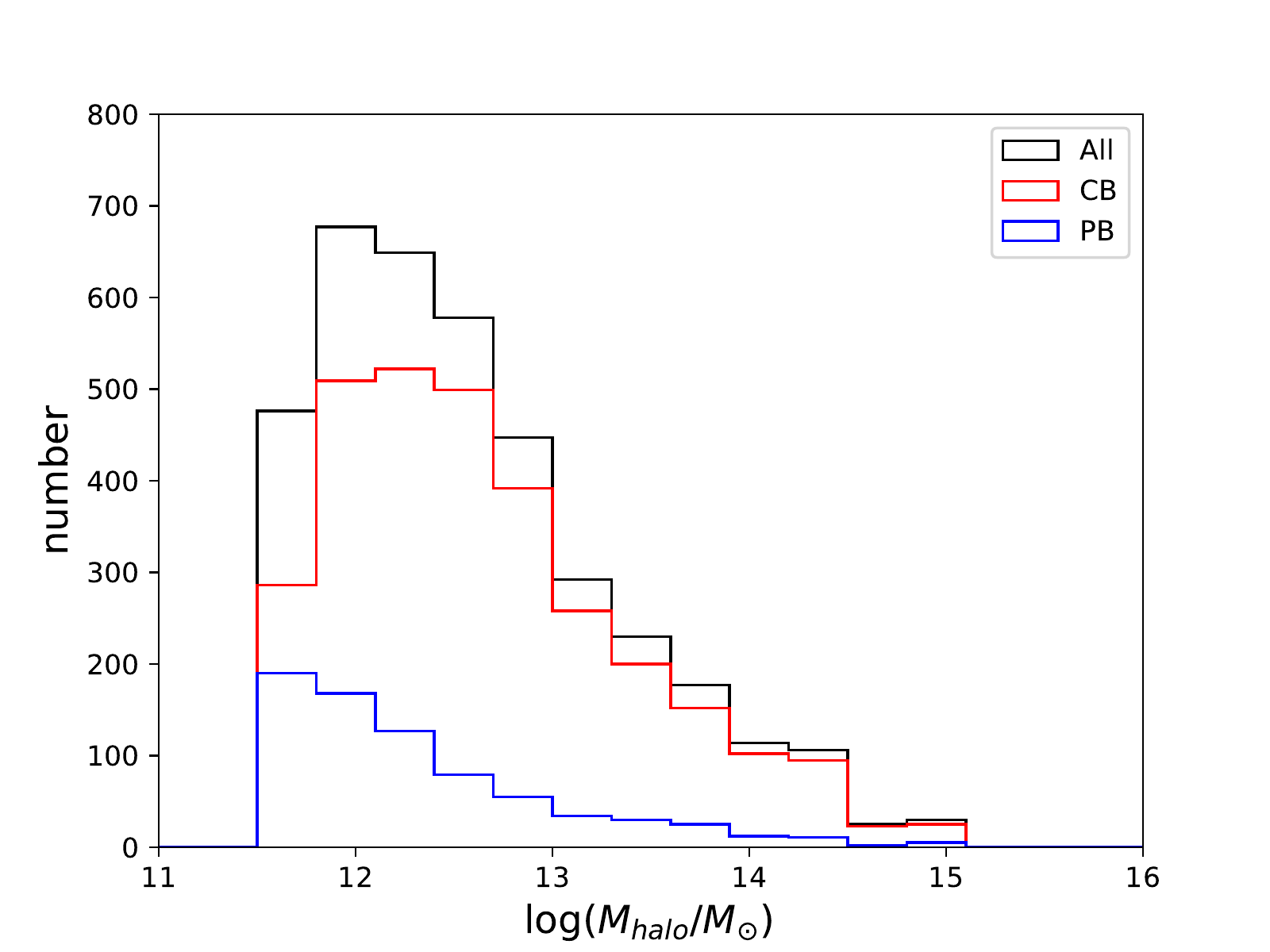} \hspace{-2.35em}
\includegraphics[width=.35\textwidth]{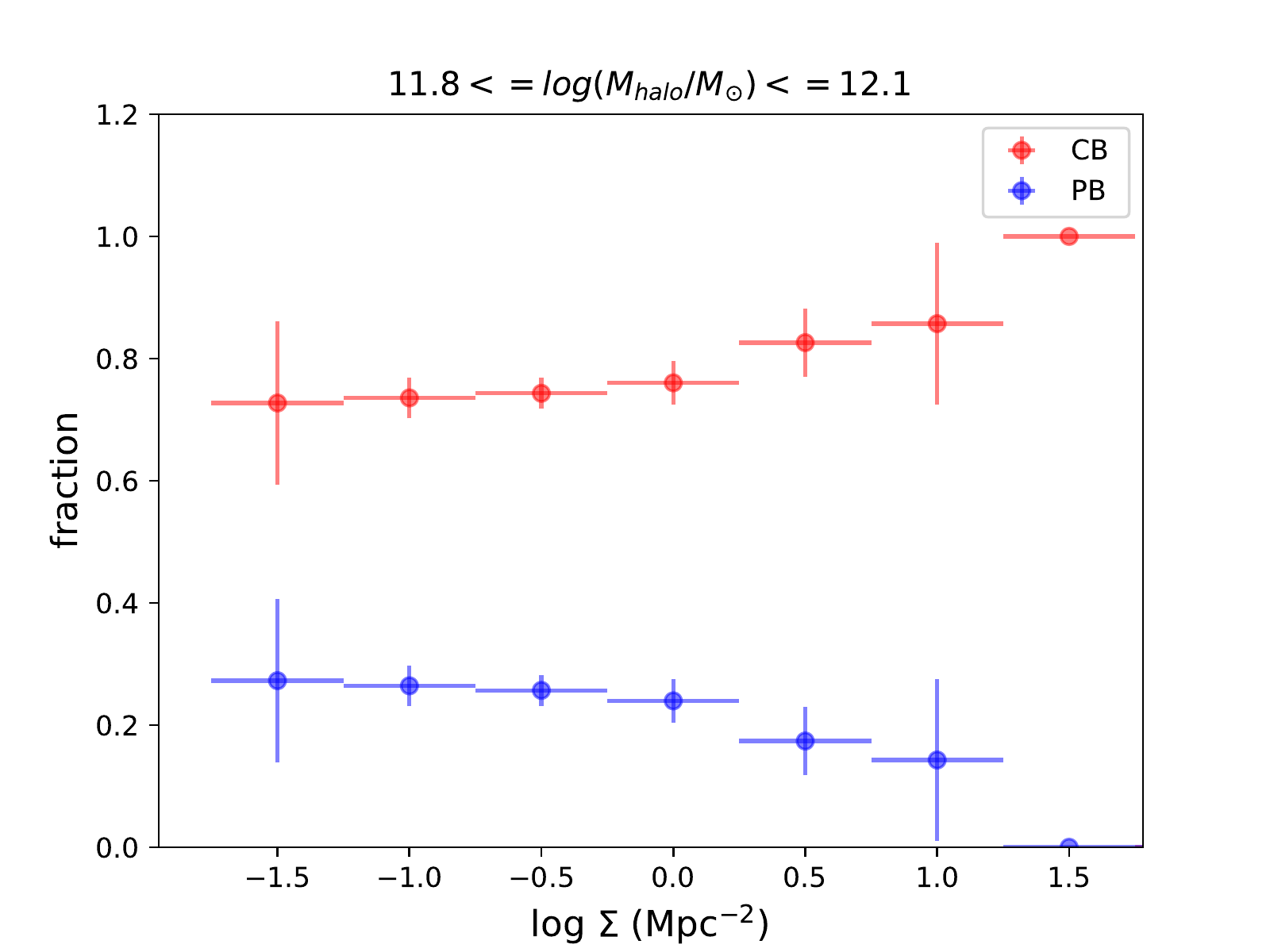}

\caption{The effect of stellar and dark matter halo mass on the environmental distribution of classical and pseudobulge host galaxies. \textbf{Left:} Scatter plot of galaxy stellar mass and the environmental density. We do not see find any strong correlation between these quantities. \textbf{Middle:} The distribution of dark matter halo mass for disc galaxies in our final sample. The red and blue histograms corresponds classical bulge hosting and pseudobulge hosting disc galaxies respectively. The black histogram is for all the disc galaxies. The halo masses are not the values of the dark matter content of the individual galaxies but rather are masses of dark matter halo of the group in which a galaxy belongs. \textbf{Right:} Plot of structural morphology density relation for galaxies when the halo mass is constrained in a narrow range of halo mass log($M_{halo}/M_{\odot}$)=[11.8,12.1]. One can notice that at fixed halo mass the structural morphology-density relation have become flat within the Poisson error bars.}
\label{fig:fig7}
\end{figure*}

We proceed by first checking the environmental dependence of stellar mass of classical and pseudobulge hosting disc galaxies in left panel of Fig. \ref{fig:fig7}. This is a scatter plot of galaxy stellar mass vs. environmental density and from this plot one does not see any significant correlation between these two quantities. It seems that correlation of bulge type with dark matter halo mass is the main driver of structural morphology density relation. Then it is interesting to check the effect on structural morphology density relation within a narrow range of dark matter halo mass. We plot the distribution of masses of host dark matter halo of galaxies in our sample in the middle panel of Fig. \ref{fig:fig7}. The distribution peaks in the bin of log($M_{halo}/M_{\odot}$)=[11.8,12.1]. There are total 678 disc galaxies in this bin with 510 (75.2\%) of them being classical bulge hosts and rest 168 (24.8\%) are pseudobulge hosts. We re-plotted the structural morphology density relation in right panel of Fig. \ref{fig:fig7}. Comparing this plot with the structural morphology density relation presented in Fig. \ref{fig:fig5}, one can see that the environmental distribution of classical and pseudobulge hosting galaxies have become flat within the error bars when halo mass is controlled. Even the value of classical bulge fraction in across the density bins of environment have changed only slightly from its global value (75.2\%). We conclude from here that the correlation of classical bulge fraction with environment is not a direct relation but is driven by  correlations between bulge types, halo mass and the environmental density.

\subsection{On quenching of classical bulge hosting disc galaxies}

We now discuss the issue of enhanced quenching of classical bulge hosting disc galaxies in dense environments. Quenching in the disc of spiral galaxies can lead to the fading of spiral arms \citep{Laurikainen2010}. The quenched spirals then evolve into S0 galaxies with time scales greater than quenching time scales \citep{Bamford2009}. However, the issue of quenching of galaxies is a complex one as there are various processes which can quench a galaxy. Moreover, the importance of these processes to quench the galaxies also depends on the environment. For example, recent studies \citep{Rizzo2018} have argued that violent fast processes like mergers and tidal interaction are not the dominant channel to form S0 galaxies in the low density environments. Internal quenching processes, for example feedback from AGN or supernovae, morphological quenching are likely to convert spiral into S0 galaxies in less dense environments consisting of small groups \citep{Mishra2018}.\par 

A high density environment opens up more avenues of quenching and morphological transformation of a spiral galaxy by action of environmental processes. Spiral galaxies entering a high density environment such as a massive galaxy clusters, can get stripped of their gas \citep{Gunn1972, Moran2007}.A massive halo hosting a spiral galaxy can also cut down the fresh supply of infalling gas onto the galaxy making it passive over time \citep{Giovanelli1986}. Tidal interaction of a spiral with other members galaxies can give rise to repetitive starburst which leads to rapid consumption of gas and quenching. The intermediate environmental regime of large galaxy groups offers different quenching avenues, with violent processes like galaxy mergers becoming important. It is known in the literature that massive mergers are more common in large galaxy groups than in clusters \citep{McIntosh2008} and are a potential formation channel of S0 galaxies \citep{Querejeta2015}.  \par

It is known that the formation of S0 galaxies happens via a number of channels depending on the property of galaxy itself and the environment in which it lives. We can get different answers to the question of how quenching leads to formation of spirals in different environments.  Much of the relevant information regarding quenching may be hidden in the starformation history of these galaxies, the understanding of which requires a detailed study of its own. In the absence of such information, we can only make speculative remarks on the role of environment as a driver of quenching. Past works \citep{Peng2010, Peng2012} have suggested that galaxy mass and the environment are the two main drivers of galaxy quenching but their effect can be separated. The mass driven quenching primarily affects the massive central galaxies having stellar mass greater than $10^{10.2} M_{\odot}$ and, is independent of the environment. On the other hand, low mass satellite galaxies are significantly affected by the environment and their quenching fraction shows strong dependence on the local environmental density. From our results in previous section and Figure \ref{fig:fig6}, one can see that most of the classical bulge hosts are massive(>$10^{10} M_{\odot}$) disc galaxies and have potential to undergo quenching driven by stellar mass. One may expect that at low environmental densities, their quenching might be happening via processes which do not strongly depend on the environments i.e. via the internal quenching mechanisms stated before. It is also likely that given a distribution of similar type of galaxies in various environments the contribution of internal process to quench the galaxies should be more or less the same. Then one can speculate that the enhanced quenching fraction in high density environments might arise due to the additional environmental effect which manifests itself by quenching the satellites galaxies, in addition to the already existing internal quenching mechanisms active in massive galaxies. \par

\subsection{General implications of the presented results}

We now highlight some of the general implications of our study. To begin
with, it should be noted that insight gained in this work on the
formation and the environmental distribution of S0 galaxies was made
possible due to the definition of morphology which is independent of
the star forming state of a galaxy. The further classification of
spirals and S0 galaxies based on the presence of bulge type pointed
towards the structural similarity between classical bulge hosting
spirals and S0 galaxies. This information coupled with the fact that
pseudobulge hosting spirals are structurally different from the
classical bulge hosting S0, led us to conclude that the classical
bulge hosting spirals are the major progenitors of the S0 galaxy
population.\par 

The results of our work further support recent works which aim to understand the origin of morphology density relation. In their work, \cite{Cappellari2011} have separated the kinematic morphology density relation of galaxies by subdividing early type galaxies into classes of fast and slow rotators (denoted by FR and SR). The fast rotators are the early type galaxies with similar stellar kinematics as spirals. They report a steady increase of fraction of FR galaxies with density at the expense of steadily decreasing spiral fraction with environmental density. From our work we find that only a certain class of spirals (classical bulge hosting-) are preferentially getting converted to S0 galaxies. Coupled with the fact that most of FR galaxies are visual S0s, our results would imply that only a certain class (and not all) of spirals might be getting transformed into FR early type galaxies.\par 

The difference seen in the size-mass relation of classical
and pseudobulge hosting disc galaxies is interesting due to a number
of reasons. For example, previous studies \citep{Wel2014, Gu2018}
concerning the evolution of galaxy size-mass relation put the star
forming and quiescent galaxies in separate categories, reporting
different relations for the two classes. In our work structural
morphology has given a different perspective to this separation of
galaxies into the star forming and quiescent classes. It tells us that
not all star forming galaxies have same size-mass relation. We
see that the classical bulge hosting spirals which are star forming,
follow a similar size-mass relation to the quenched classical bulge
hosting S0 galaxies. It is also interesting to note that at similar
stellar masses the pseudobulge hosting spirals are larger in size as
compared to classical bulge hosting spirals. This means that, on
average, the global mass distribution is more sparse in pseudobulge
hosting galaxies as compared to the ones hosting a classical bulge. In
the standard picture of galaxy formation the size of the galaxy disc
is linked to the size of the dark matter halo and the fraction of halo
angular momentum retained by the baryons \citep{Mo2010}. Therefore it
may be possible that differences seen in the size-mass relation of
classical and pseudobulge hosting disc galaxies are due to the
different distribution of underlying dark matter and, the different
angular momentum content of these galaxies. However, in this paper, we have not tested these hypotheses and is a subject matter of future interest.

\subsection{Limitations of this work}

We now state some of the limitations of our work which do not affect
the main results of this paper but are important nonetheless. In our
work, we have neglected the pseudobulge hosting S0 galaxies primarily
because they constitute a small fraction of the S0 population and do not
show any significant trend with environment, perhaps again due to
small number statistics, with environmental density. At some places we
also have clubbed them together with pseudobulge hosting spirals for
clarity of presentation even though they follow a different size-mass
relation. The pseudobulge hosting S0 galaxies are interesting in their
own right as they have size-mass relation intermediate to classical
bulge hosting disc galaxies and pseudobulge hosting spirals. They are
also knows to show peculiar behaviour in terms of their stellar
population properties. In our previous work \citep{Mishra2017b}, we
have seen that age distribution of pseudobulges of S0 galaxies is
bimodal, while no such bimodality is seen in pseudobulges of
spirals. Also, as can be seen in the $NUV-r$ colour mass diagram a
significant fraction of the pseudobulge hosting galaxies are star
forming, something which is an aberration from the classical concept
of red and dead S0 galaxies. Pseudobulge hosting S0 galaxies seems to
have undergone complex evolutionary paths and perhaps a detailed
modelling of star formation history will help us to understand these
objects better. We would also like to mention that while discussing
the effect of environment on quenching, we have not taken the
preprocessing factor into account \citep{Wetzel2013}. It is true that
environmental effects are usually used to explain the quenching of
galaxies in dense environments like clusters but these dense regions are known to evolve by accretion of smaller systems like groups of galaxies. Therefore, it is possible that the quenching of galaxies has already taken place in the group stage and assembly into dense
environment came later \citep{Wetzel2013, Bianconi2018}. Separating the galaxies in denser environments into those which are preprocessed and those which are quenched
actually in dense environment will require a detailed modelling of
star formation history of galaxies. Last but not the least, we would
like to remind our reader that all disc galaxies in our sample do
not host a bar. The exclusion of barred galaxies from our sample was
due the unavailability of a bulge + bar + disc decomposition for our
sample. Fitting a barred galaxy with only a bulge + disc model might
have led to erroneous classification of bulge types. All the result
presented in our study should be viewed keeping these limitations in mind.

\section{Summary and conclusion}

\label{sec:sum}

In this work we have attempted to understand why S0 galaxies commonly occur in high density environments. Constructing a sample of 2541 spirals and 2032 S0 SDSS galaxies, we have divided each visual morphological class into two structural morphology subclasses of classical and pseudobulge hosting galaxies. We have studied their environmental distribution coupled with information on structural and starformation properties. The main results of our work can be summarised as follows: \\

i) We find that classical and pseudobulge fractions (60\% and 40\% respectively) are comparable in case of spirals although almost all ($\sim$93\%) S0 galaxies are classical bulge hosts. We also find that these fractions change as a function of environmental density. The classical bulge hosting spirals and S0 galaxies follow the same initial increasing trend as one moves from low to mid environmental density regime. But as one keep on moving towards high densities the fraction of classical bulge hosting spiral takes a downward turn and this is accompanied by a steady increase in fraction of S0 galaxies. The fraction of pseudobulge hosting spirals, on the other hand, steadily decreases with increase in environmental density. \par

ii) From the viewpoint of size-stellar mass relation, the classical bulge hosting spirals and S0s are found to be structurally similar. The pseudobulge hosting spirals show significantly different distribution on the size-mass plane. Inspecting the global starformation properties of galaxies on the $NUV-r$ colour-stellar mass diagram, we find that almost all of the pseudobulge hosting spiral galaxies are starforming. However, a significant fraction of classical bulge hosting spirals are situated either in the green valley or in the quenched sequence. From these results, we argue that most of the present day S0 galaxies are formed via morphological transformation of classical bulge hosting spiral galaxies. \par

iii) We find that there exists a biased environmental distribution of galaxy structural morphology, such that the fraction of classical bulge host spiral galaxies increases with increase in environmental density. Since classical bulge hosting spirals are the main progenitors of S0 galaxies, it is not surprising that one finds more S0 galaxies in high density environments. The observed environmental trend of classical bulge hosting galaxies with environment holds even when all disc galaxies are taken together, irrespective of their visual types.\par

iv) In addition to the above mentioned biased environmental distribution, we also explore the role  of environment on the quenching of classical bulge hosting spirals. We separate the classical bulge hosting spiral galaxies into starforming and quenched classes, and examine their trend with the environment. We find that at denser environments, more quenched classical bulge hosting spirals are forming at the expense of star forming classical bulge hosting spiral galaxies.  \par

v) There exists a correlation between the structural morphology and the mass of the parent dark matter halo of galaxies, such that the dark matter haloes more massive than $10^{13} M_{\odot}$ prefer to host classical bulge hosting disc galaxies. One finds very few pseudobulge hosting disc galaxies in haloes having mass greater than than $10^{13} M_{\odot}$. When the mass of the dark matter halo is controlled, the environmental distribution of classical bulge hosting disc galaxies becomes significantly flat. It is likely that the environmental distribution of classical bulge hosting galaxies, irrespective of the morphological classes, is driven by its correlation with the mass of the dark matter halo. \par

We conclude that the major reason for the frequent occurrence of S0 galaxies in denser environments is due to the existence of an already biased environmental distribution of progenitors of S0 galaxies, combined with the fact that environment is more efficient in quenching disc galaxies in high density environments as compared to low density environments. Further work on the modelling of star formation history of classical bulge hosting spirals and S0 galaxies in different environments will provide better insight on the quenching mechanisms responsible for the formation of S0 galaxies.

\section*{Acknowledgements}

We thank the anonymous referee whose insightful comments have improved both the content and presentation of this paper.

%%%%%%%%%%%%%%%%%%%%%%%%%%%%%%%%%%%%%%%%%%%%%%%%%%



%\bibliographystyle{mnras}


%%%%%%%%%%%%%%%%%%%%%%%%%%%%%%%%%%%%%%%%%%%%%%%%%%

%%%%%%%%%%%%%%%%% APPENDICES %%%%%%%%%%%%%%%%%%%%%

%%%%%%%%%%%%%%%%%%%%%%%%%%%%%%%%%%%%%%%%%%%%%%%%%%

% Don't change these lines
\bsp	% typesetting comment
\label{lastpage}
\end{document}